\documentclass[%
print,
 amsmath,amssymb,
 aps,
]{revtex4}

\usepackage{times}
\usepackage{graphicx}
\usepackage{dcolumn}
\usepackage{bm}
\usepackage{dsfont}
\usepackage{mathrsfs}
\usepackage[landscape,papersize={297.1mm,210mm},left=1.9cm,right=1.6cm,top=2.7cm,bottom=2.8cm]{geometry}
\usepackage[                   
            pdfstartview=FitH,
            colorlinks, 
            pdfborder=001,   
            linkcolor=blue,
            anchorcolor=blue,
            citecolor=blue,
            urlcolor=blue
            ]{hyperref}
\usepackage{amsthm}
\makeatletter

\newcommand{\Rmnum}[1]{\expandafter\@slowromancap\romannumeral #1@}

\makeatother

\begin{document}
\title{The fidelity of controlled quantum teleportation in a noisy environment}

\author{Wen-Jing Wei$^{1}$}
\author{Feng-Li Yan$^{1}$}
\email{flyan@hebtu.edu.cn}
\author{Ting Gao$^{2}$}
\email{gaoting@hebtu.edu.cn}

\affiliation {$^1$ College of Physics, Hebei Key Laboratory of Photophysics Research and Application, Hebei Normal University, Shijiazhuang 050024, China\\
$^2$ School of Mathematical Sciences, Hebei Normal University, Shijiazhuang 050024, China}

\begin{abstract}
In this work, we investigate controlled quantum teleportation in the presence of noisy channels acting on the three-qubit resource state. We employ a series of generalized noisy channels that bridge the dephasing channels and amplitude damping channels while encompassing extensive intermediate scenarios. We provide an in-depth analysis of the degradation of the maximal average fidelity and the optimal average fidelity in controlled quantum teleportation induced by such noisy channels by deriving the analytical expression and examining several special cases. The analytical expression shows that attaining the optimal average fidelity requires Charlie's cooperation in performing a measurement at suitably chosen angles, and is also related to the initial state and the channel parameters. Our analysis reveals that the optimal average fidelity does not always decrease monotonically with the evolution parameter, instead, it first decreases and then increases. This non-monotonic behavior depends on the entanglement of the initial resource state, as well as on the parameters of the channel traversed by the first qubit.\\

\end{abstract}


\maketitle

\section{Introduction}

Quantum teleportation is a method of transmitting unknown quantum states between the sender Alice and the distant receiver Bob, and is regarded as one of the most essential application of  quantum entanglement resources. In the pioneering work on standard quantum teleportation by Bennett et al. \cite{PRL70.1895}, it is assumed that Alice and Bob share previously an entangled Einstein-Podolsky-Rosen pair. Then the sender Alice makes a Bell state measurement \cite{PRA59.3259} on the qubit and one particle of entangled pair and informs the receiver Bob the result of the  measurement via a classical information channel.  Based on the sender Alice's  measurement result Bob performs  a corresponding unitary operation on the other particle of the EPR pair,   then the  unknown quantum state of the qubit at Alice's place has  been transmitted   to the receiver Bob perfectly. As quantum teleportation has advanced, it has been studied and generalized to numerous settings \cite{PRA68.050302,PRA67.032302,PRA68.022321,PRA66.042326,PRA49.1473,PRL80.869,PRA64.064304,PRA73.042309,PRA70.022329,PRA61.022308,PLA305.12,JOB6.S844,PLA316.297,PRA72.022338,PRA102.012414,PRA108.042620,PRR6.033313,PRL127.080502}.

By introducing  a controller Charlie in the standard quantum teleportation, the controlled quantum teleportation \cite{PRA58.4394} was established,  where the unknown quantum state can be recovered  only if  the controller Charlie consents to cooperate. In the seminal work of Karlsson and Bourennane \cite{PRA58.4394}, they exploited the entanglement property of the Greenberger-Horene-Zeilinger state to teleport an unknown quantum state via local operations and classical communication  under the supervision of the controller Charlie. There has been extensive researches on controlled quantum teleportation scheme \cite{PRA70.022329,PRA61.022308,PLA316.297,PRA72.022338}. Gao et al. \cite{EPL84.50001},  investigated optimal controlled teleportation and obtained the exact tripartite entangled state  which   can be used for perfect controlled teleportation. A number of quantum communication protocols, including the construction of quantum network \cite{PRR6.L032062} and long-distance communication \cite{PRR3.023038}, rely on the ability to teleport quantum states under controlled conditions. Controlled quantum teleportation addresses this need by providing a verifiable and secure mechanism, with potential extensions to quantum repeaters and high-efficiency quantum information processing \cite{PRB94.235446}.

Noisy quantum teleportation refers to the remote transfer of unknown quantum states via quantum entanglement and classical communication in realistic physical systems that are subject to environmental noise and decoherence. First proposed by Bennett et al. \cite{PRL76.722}, this scheme has since been investigated in a considerable body of work \cite{PRA66.022316,EPJ9,NPJ7,PRA62.012311,PRA65.022302,PRA92.012338,PRA100.062311,PRA90.042332,PRA109.052606}.  From these works, a straightforward conclusion follows: there exist channels acting on certain pure states such that the resulting noise is minimally detrimental to the fidelity of quantum teleportation. Evidently, the fidelity is an important quality factor of the teleportation,  which serves as a measure of teleportation success, and a reflection of  the real physical environmental noise. Despite some progress in noisy quantum teleportation, research on controlled quantum teleportation involving three-qubit states \cite{PRL85.1560} as a quantum entanglement resource   in the  noisy channels is still relatively limited.

In this paper, we investigate the influence of noisy channels on controlled quantum teleportation under general conditions. We use a tripartite entangled state \cite{EPL84.50001},  which  can be used for perfect controlled teleportation as the entanglement resource.  We analysis  the fidelity of the  controlled teleportation in a series of generalized noisy channels \cite{PRA97.052332} that bridge the dephasing channels and amplitude damping channels while encompassing extensive intermediate scenarios. Such an analysis renders the study of controlled quantum teleportation over noisy channels more comprehensive. Furthermore, we give the analytic expression for the maximal average fidelity.

The content of this paper is arranged as follows. In Sec. \ref{Q2}, we describe the effect of noisy channels on controlled quantum teleportation, the maximal average fidelity, and the Kraus operators of the quantum channels acting on the initial resource state. In Sec. \ref{Q3}, we calculate the maximal average fidelity and the optimal average fidelity of controlled quantum teleportation via noisy channels and derive their analytical expressions. To illustrate the concrete influence of  noisy channel on the the maximal average fidelity of controlled quantum teleportation, in the following section, we enumerate some special cases including the standard controlled quantum teleportation, a single qubit passing through a noisy channel,  two qubits passing through noisy channels as well as three qubits passing through the identical channel. A summary is given in Sec. \ref{Q8}.

\section{The maximal average fidelity  via the noisy quantum channels}\label{Q2}

Gao et al. \cite{EPL84.50001}, demonstrated  that  a tripartite  state can be utilized  for perfect   controlled teleportation of an unknown qubit state $|\psi\rangle=a|0\rangle+b|1\rangle$ with  unit probability and with unit fidelity if  and only if it is  equivalent to
\begin{equation}\label{1}
|\Psi\rangle_{ABC}=a_{0}|000\rangle+a_{1}|100\rangle+\frac{1}{\sqrt{2}}|111\rangle
\end{equation}
under local operation and classical communications.  Here $a_{0}>0, ~a_{1}\geq0, ~~a_{0}^{2}+a_{1}^{2}=\frac{1}{2}; ~~|a|^{2}+|b|^{2}=1$,  and the qubit $A$ belongs to the controller Charlie, qubit $B$ belongs to the sender Alice, while Bob has qubit $C$.

Suppose that Alice,   Bob  and  Charlie establish  previously a tripartite entangled state (\ref{1}).  The controller Charlie measures his qubit $A$ with Positive Operator-Valued Measurement $M=\{M_{1},M_{2}\}$,
\begin{equation}
\begin{aligned}
&M_{1}=(\cos\frac{\theta}{2}|0\rangle + {\rm e}^{{\rm i}\varphi}\sin\frac{\theta}{2}|1\rangle)(\cos\frac{\theta}{2}\langle0|+ {\rm e}^{-{\rm i}\varphi}\sin\frac{\theta}{2}\langle1|),\\
&M_{2}= (\sin\frac{\theta}{2}|0\rangle-{\rm e}^{{\rm i}\varphi}\cos\frac{\theta}{2}|1\rangle)(\sin\frac{\theta}{2}\langle0|- {\rm
e}^{-{\rm i}\varphi}\cos\frac{\theta}{2}\langle1|),\\
&M_{1}+M_{2}=I,
\end{aligned}
\end{equation}
where $\theta\in [0,\pi]$, $\varphi\in [0,2\pi]$. If  Charlie consents to teleport the unknown quantum state between  Alice and  Bob,   then Charlie sends his measurement results to Alice and Bob via a classical communication channel. According to the measurement result of Charlie, Alice  and Bob  can  deliver the unknown quantum $|\psi\rangle=a|0\rangle+b|1\rangle$ to Bob by the protocol of standard teleportation. However, if the controller Charlie does not inform Alice and Bob his measurement outcome, the teleportation can not been completed. Thus Charlie can control  the teleportation from the sender Alice to the receiver Bob.

However,   the controlled teleportation are tempered  by the fact that the real systems  suffer from the  unwanted interaction with  the  environment. This interaction is called  quantum noisy channel and  described  by Kraus operators \cite{Nielsen}. Assume that  the initially resource   state $|\Psi\rangle_{ABC}$  passes through a quantum channel $\Lambda_{ABC}$ represented by a set of Kraus operators $\{\Pi_{\mu}\}$, the channel converts $|\Psi\rangle\langle\Psi|$ into a valid mixed resource state
\begin{equation}
\rho_{ABC}=\Lambda_{ABC}(|\Psi\rangle\langle\Psi|)=\sum\limits_\mu\Pi_{\mu}|\Psi\rangle\langle\Psi|\Pi_{\mu}^{\dag}.
\end{equation}
Since Alice, Bob, and Charlie are usually separated, we focus on the case where particles $A, B,$ and $C$ pass through local channels independently. Therefore
\begin{equation}
 \Lambda_{ABC}=\Lambda_{A}\otimes\Lambda_{B}\otimes\Lambda_{C}, ~~~\Pi_{\mu}\rightarrow\Pi_{\alpha\beta\gamma}=Q_{\alpha}\otimes K_{\beta}\otimes R_{\gamma},
\end{equation}
where $\{Q_{\alpha}\}$, $\{K_{\beta}\}$, and $\{R_{\gamma}\}$ stand for the sets of Kraus operators associated to $\Lambda_{A}$, $\Lambda_{B}$, and $\Lambda_{C}$, respectively. Then the mixed resource state $\rho_{ABC}$ can be written as
\begin{equation}
\rho_{ABC}=\sum\limits_{\alpha\beta\gamma} Q_{\alpha}\otimes K_{\beta}\otimes R_{\gamma}|\Psi\rangle\langle\Psi|Q_{\alpha}^{\dagger}\otimes K_{\beta}^{\dagger}\otimes R_{\gamma}^{\dagger}.
\end{equation}

In the real physical process of the controlled  teleportation,  after the controller Charlie performs the measurements $M=\{M_{1},M_{2}\}$ on the qubit $A$, the quantum resource states become
\begin{equation}
\rho_{1ABC}=(M_{1}\otimes I\otimes I)\rho_{ABC}(M_{1}^{\dagger}\otimes I \otimes I)/{\rm Tr}[(M_{1}\otimes I\otimes I)\rho_{ABC}(M_{1}^{\dagger}\otimes I \otimes I)]
\end{equation}
and
\begin{equation}
\rho_{2ABC}=(M_{2}\otimes I\otimes I)\rho_{ABC}(M_{2}^{\dagger}\otimes I \otimes I)/{\rm Tr}[(M_{2}\otimes I\otimes I)\rho_{ABC}(M_{2}^{\dagger}\otimes I \otimes I)]
\end{equation}
with probabilities
\begin{equation}
\begin{aligned}
&p_{1}={\rm Tr}[(M_{1}\otimes I\otimes I)(\sum\limits_{\alpha\beta\gamma} Q_{\alpha}\otimes K_{\beta}\otimes R_{\gamma}|\Psi\rangle\langle\Psi|Q_{\alpha}^{\dagger}\otimes K_{\beta}^{\dagger}\otimes R_{\gamma}^{\dagger})(M_{1}^{\dagger}\otimes I\otimes I)]\\
\end{aligned}
\end{equation}
and
\begin{equation}
\begin{aligned}
&p_{2}={\rm Tr}[(M_{2}\otimes I\otimes I)(\sum\limits_{\alpha\beta\gamma} Q_{\alpha}\otimes K_{\beta}\otimes R_{\gamma}|\Psi\rangle\langle\Psi|Q_{\alpha}^{\dagger}\otimes K_{\beta}^{\dagger}\otimes R_{\gamma}^{\dagger})(M_{2}^{\dagger}\otimes I\otimes I)]
\end{aligned}
\end{equation}
respectively. Here $I$ is an identity operator.

If the controller Charlie agrees to  proceed the teleportation, he informs Alice and Bob his measurement outcome. Then Alice and Bob can realize the teleportation by mixed resource states
$\rho_{1BC}={\rm Tr}_A[{\rho_{1ABC}}]$ or $\rho_{2BC}={\rm Tr}_A[{\rho_{2ABC}}]$ with certain fidelity of the teleportation. It has been proved that  with a mixed resource state $\rho_{BC}$ Alice and Bob  can achieve the maximal  fidelity of the teleportation \cite{PRA60.1888}
\begin{equation}
F_{\rm max}=\max\limits_{i\in \{1,2,3,4\}} \{\frac{1}{3}+\frac{2}{3}\langle\Phi_{i}|\rho_{BC}|\Phi_{i}\rangle\}],
\end{equation}
where $\Phi_{1}=\frac {1}{\sqrt{2}}(|00\rangle +|11\rangle)$, $\Phi_{2}=\frac {1}{\sqrt{2}}(|00\rangle -|11\rangle)$, $\Phi_{3}=\frac {1}{\sqrt{2}}(|01\rangle +|10\rangle)$, $\Phi_{4}=\frac {1}{\sqrt{2}}(|01\rangle -|10\rangle)$  are four Bell states.

In  the noisy quantum environment,  the maximal average fidelity of controlled teleportation is defined as
\begin{equation}
\begin{aligned}
~~&F_{\rm max}\\
=&p_{1}F_{\rm 1max}+p_{2}F_{\rm2max}\\
=&p_{1}[\frac{1}{3}+\frac{2}{3}\max\limits_{i\in \{1,2,3,4\}} \{\langle\Phi_{i}|\rho_{1BC}|\Phi_{i}\rangle\}]+p_{2}[\frac{1}{3}+\frac{2}{3}\max\limits_{j\in \{1,2,3,4\}} \{\langle\Phi_{j}|\rho_{2BC}|\Phi_{j}\rangle\}],
\end{aligned}
\end{equation}
where
\begin{equation}
\begin{aligned}
&F_{\rm 1max}=\frac{1}{3}+\frac{2}{3}\max\limits_{i\in \{1,2,3,4\}}\{\langle\Phi_{i}|\frac{{\rm Tr_{A}}[(M_{1}\otimes I\otimes I)(\sum\limits_{\alpha\beta\gamma} Q_{\alpha}\otimes K_{\beta}\otimes R_{\gamma}|\Psi\rangle\langle\Psi|Q_{\alpha}^{\dagger}\otimes K_{\beta}^{\dagger}\otimes R_{\gamma}^{\dagger})(M_{1}^{\dagger}\otimes I\otimes I)]}{{{\rm Tr}[(M_{1}\otimes I\otimes I)(\sum\limits_{\alpha\beta\gamma} Q_{\alpha}\otimes K_{\beta}\otimes R_{\gamma}|\Psi\rangle\langle\Psi|Q_{\alpha}^{\dagger}\otimes K_{\beta}^{\dagger}\otimes R_{\gamma}^{\dagger})(M_{1}^{\dagger}\otimes I\otimes I)]}}|\Phi_{i}\rangle\},\\
&F_{\rm 2max}=\frac{1}{3}+\frac{2}{3}\max\limits_{j\in \{1,2,3,4\}}\{\langle\Phi_{j}|\frac{{\rm Tr_{A}}[(M_{2}\otimes I\otimes I)(\sum\limits_{\alpha\beta\gamma} Q_{\alpha}\otimes K_{\beta}\otimes R_{\gamma}|\Psi\rangle\langle\Psi|Q_{\alpha}^{\dagger}\otimes K_{\beta}^{\dagger}\otimes R_{\gamma}^{\dagger})(M_{2}^{\dagger}\otimes I\otimes I)]}{{{\rm Tr}[(M_{2}\otimes I\otimes I)(\sum\limits_{\alpha\beta\gamma} Q_{\alpha}\otimes K_{\beta}\otimes R_{\gamma}|\Psi\rangle\langle\Psi|Q_{\alpha}^{\dagger}\otimes K_{\beta}^{\dagger}\otimes R_{\gamma}^{\dagger})(M_{2}^{\dagger}\otimes I\otimes I)]}}|\Phi_{j}\rangle\}.
\end{aligned}
\end{equation}


\section{The Fidelity through Three Noisy Channels}\label{Q3}

Let us  consider the case where qubits $A, B$ and $C$ all pass through noisy channels, which results in the emergence of the mixed resource state $\rho_{ABC}$ before the measurement phase of the protocol. This scenario corresponds to $|\Psi\rangle\langle\Psi|$ passing through a channel $\Lambda_{ABC}=\Lambda_{A}\otimes \Lambda_{B}\otimes \Lambda_{C}$, where $\Lambda_{A}$, $\Lambda_{B}$ and $\Lambda_{C}$ are generalized noisy channels. Without loss of generality we choose  the corresponding Kraus operators \cite{PRA97.052332} as
\begin{equation}\label{Q}
\begin{aligned}
Q_{0}=\begin{bmatrix} 1 &   0\\ \mathrm0 & \sqrt{1-q}  \end{bmatrix},~~
Q_{1}=\sqrt{q}\begin{bmatrix} 0 &   \cos\zeta_{A}\\ \mathrm0 & \sin\zeta_{A}  \end{bmatrix},\\
K_{0}=\begin{bmatrix} 1 &   0\\ \mathrm0 & \sqrt{1-k}  \end{bmatrix},~~
K_{1}=\sqrt{k}\begin{bmatrix} 0 &   \cos\zeta_{B}\\ \mathrm0 & \sin\zeta_{B}  \end{bmatrix},\\
R_{0}=\begin{bmatrix} 1 &   0\\ \mathrm0 & \sqrt{1-r}  \end{bmatrix},~~
R_{1}=\sqrt{r}\begin{bmatrix} 0 &   \cos\zeta_{C}\\ \mathrm0 & \sin\zeta_{C}  \end{bmatrix},\\
\end{aligned}
\end{equation}
where $\zeta_{A}, \zeta_{B}, \zeta_{C}\in[0,2\pi]$ and $q, k, r \in [0,1]$ are called channel parameters and  evolution parameters respectively, both of them  are determined by the noisy channel.


 When $\zeta_A ~(\zeta_B,~\zeta_C )=0$, Eq.(\ref{Q}) becomes the Kraus operators of the amplitude damping channel, and when $\zeta_A~(\zeta_B,~\zeta_C )=\frac{\pi}{2}$, it describes  a dephasing channel. Both are typical decoherence channels, so the general channel in Eq.(\ref{Q}) is called the generalized noisy channel.

Thus the maximal average fidelity through the three noisy channels described by Eq.(\ref{Q}) reads
\begin{equation}
\begin{aligned}
~~&F_{\max}\\
=&p_{1}F_{1\max}+p_{2}F_{2\max}\\
=&p_{1}[\frac{1}{3}+\frac{2}{3p_{1}}\max\limits_{i\in \{1,2,3,4\}}\{\langle\Phi_{i}|{\rm Tr}_{A}[(M_{1}\otimes I\otimes I)(\sum\limits_{\alpha\beta\gamma} Q_{\alpha}\otimes K_{\beta}\otimes R_{\gamma}|\Psi\rangle\langle\Psi|Q_{\alpha}^{\dagger}\otimes K_{\beta}^{\dagger}\otimes R_{\gamma}^{\dagger})(M_{1}^{\dagger}\otimes I\otimes I)]|\Phi_{i}\rangle\}]\\
&+p_{2}[\frac{1}{3}+\frac{2}{3p_{2}}\max\limits_{j\in \{1,2,3,4\}}\{\langle\Phi_{j}|{\rm Tr}_{A}[(M_{2}\otimes I\otimes I)(\sum\limits_{\alpha\beta\gamma} Q_{\alpha}\otimes K_{\beta}\otimes R_{\gamma}|\Psi\rangle\langle\Psi|Q_{\alpha}^{\dagger}\otimes K_{\beta}^{\dagger}\otimes R_{\gamma}^{\dagger})(M_{2}^{\dagger}\otimes I\otimes I)]|\Phi_{j}\rangle\}]\\
=&{\max\limits_{{i,j}\in \{1,2,3,4\}}}\{\frac{1}{3}+\frac{2}{3}\langle\Phi_{i}|{\rm Tr}_{A}[(M_{1}\otimes I\otimes I)(\sum\limits_{\alpha\beta\gamma} Q_{\alpha}\otimes K_{\beta}\otimes R_{\gamma}|\Psi\rangle\langle\Psi|Q_{\alpha}^{\dagger}\otimes K_{\beta}^{\dagger}\otimes R_{\gamma}^{\dagger})(M_{1}^{\dagger}\otimes I\otimes I)]|\Phi_{i}\rangle\\
&+\frac{2}{3}\langle\Phi_{j}|{\rm Tr}_{A}[(M_{2}\otimes I\otimes I)(\sum\limits_{\alpha\beta\gamma} Q_{\alpha}\otimes K_{\beta}\otimes R_{\gamma}|\Psi\rangle\langle\Psi|Q_{\alpha}^{\dagger}\otimes K_{\beta}^{\dagger}\otimes R_{\gamma}^{\dagger})(M_{2}^{\dagger}\otimes I\otimes I)]|\Phi_{j}\rangle\},
\end{aligned}
\end{equation}
where
\begin{equation}
\begin{aligned}
p_{1}=\sin^{2}\frac{\theta}{2}+a_{0}^{2}\cos\theta+a_{0}a_{1}\sqrt{1-q}\sin\theta\cos\varphi+(1-a_{0}^{2})q\cos^{2}\zeta_{A}\cos\theta+\frac{1}{2}(1-a_{0}^{2})q\sin2\zeta_{A}\sin\theta\cos\varphi,\\
p_{2}=\cos^{2}\frac{\theta}{2}-a_{0}^{2}\cos\theta-a_{0}a_{1}\sqrt{1-q}\sin\theta\cos\varphi-(1-a_{0}^{2})q\cos^{2}\zeta_{A}\cos\theta-\frac{1}{2}(1-a_{0}^{2})q\sin2\zeta_{A}\sin\theta\cos\varphi.\\
\end{aligned}
\end{equation}

 For  convenience we introduce the following parameters
\begin{equation}\label{xyz}
x=\sqrt{2}a_{1}\sqrt{(1-k)(1-r)}\geq 0, ~~~~~y=\frac {1}{4}kr\sin2\zeta_{B}\sin2\zeta_{C}, ~~~~z=\frac{1}{2}k\cos^{2}\zeta_{B}+\frac{1}{2}r\cos^{2}\zeta_{C}-kr\cos^{2}\zeta_{B}\cos^{2}\zeta_{C}>0.
\end{equation}

A tedious calculation shows that
\begin{equation}\label{188}
F_{\max}=\max\{\frac{2}{3}, F_{\max}^{1},F_{\max}^{2},\cdots, F_{\max}^{16} \},
\end{equation}
where $\frac {2}{3}$ is the classical threshold value which can be achieved  without any entanglement resource \cite{PRA60.1888,PRL74.1259}, and
\begin{equation}\nonumber
\begin{aligned}
F_{\max}^{1}=&\frac{2}{3}+\frac{1}{3}(x+y-z),\\
F_{\max}^{2}=&\frac{2-z}{3}+\frac{1}{3}[(2q\cos^{2}\zeta_{A}-1)(x+y)\cos\theta+(\frac{a_{0}}{a_{1}}\sqrt{1-q}x+q(x+y)\sin2\zeta_{A})\sin\theta\cos\varphi],\\
F_{\max}^{3}=&\frac{2}{3}+\frac{1}{3}\{[a_{0}^{2}(1-q\cos^{2}\zeta_{A})+(q\cos^{2}\zeta_{A}-\frac{1}{2})(1+x-2z)]\cos\theta+[a_{0}\sqrt{1-q}(a_{1}+\frac{x}{2a_{1}})\\
&+q\sin2\zeta_{A}(\frac{1}{2}+\frac{1}{2}x-z-\frac{1}{2}a_{0}^{2})]\sin\theta\cos\varphi+\frac{1}{2}x+y-\frac{1}{2}\},\\
F_{\max}^{4}=&\frac{2}{3}+\frac{1}{3}\{[a_{0}^{2}(1-q\cos^{2}\zeta_{A})+(q\cos^{2}\zeta_{A}-\frac{1}{2})(1+x-2z+2y)]\cos\theta+[a_{0}\sqrt{1-q}(a_{1}+\frac{x}{2a_{1}})\\
&+q\sin2\zeta_{A}(\frac{1}{2}+\frac{1}{2}x-z-\frac{1}{2}a_{0}^{2}+y)]\sin\theta\cos\varphi+\frac{1}{2}x-\frac{1}{2}\},\\
F_{\max}^{5}=&\frac{2-z}{3}+\frac{1}{3}[(1-2q\cos^{2}\zeta_{A})(x+y)\cos\theta-(\frac{a_{0}}{a_{1}}\sqrt{1-q}x+q(x+y)\sin2\zeta_{A})\sin\theta\cos\varphi],\\
F_{\max}^{6}=&\frac{2}{3}-\frac{1}{3}(x+y+z),\\
F_{\max}^{7}=&\frac{2}{3}+\frac{1}{3}\{[a_{0}^{2}(1-q\cos^{2}\zeta_{A})+(q\cos^{2}\zeta_{A}-\frac{1}{2})(1-x-2z-2y)]\cos\theta+[a_{0}\sqrt{1-q}(a_{1}-\frac{x}{2a_{1}})\\
&+q\sin2\zeta_{A}(\frac{1}{2}-\frac{1}{2}x-z-\frac{1}{2}a_{0}^{2}-y)]\sin\theta\cos\varphi-\frac{1}{2}x-\frac{1}{2}\},\\
\end{aligned}
\end{equation}
\begin{equation}\label{189}
\begin{aligned}
F_{\max}^{8}=&\frac{2}{3}+\frac{1}{3}\{[a_{0}^{2}(1-q\cos^{2}\zeta_{A})+(q\cos^{2}\zeta_{A}-\frac{1}{2})(1-x-2z)]\cos\theta+[a_{0}\sqrt{1-q}(a_{1}-\frac{x}{2a_{1}})\\
&+q\sin2\zeta_{A}(\frac{1}{2}-\frac{1}{2}x-z-\frac{1}{2}a_{0}^{2})]\sin\theta\cos\varphi-\frac{1}{2}x-y-\frac{1}{2}\},\\
F_{\max}^{9}=&\frac{2}{3}+\frac{1}{3}\{[-a_{0}^{2}(1-q\cos^{2}\zeta_{A})+(\frac{1}{2}-q\cos^{2}\zeta_{A})(1+x-2z)]\cos\theta+[-a_{0}\sqrt{1-q}(a_{1}+\frac{x}{2a_{1}})\\
&-q\sin2\zeta_{A}(\frac{1}{2}+\frac{1}{2}x-z-\frac{1}{2}a_{0}^{2})]\sin\theta\cos\varphi+\frac{1}{2}x+y-\frac{1}{2}\},\\
F_{\max}^{10}=&\frac{2}{3}+\frac{1}{3}\{[-a_{0}^{2}(1-q\cos^{2}\zeta_{A})+(\frac{1}{2}-q\cos^{2}\zeta_{A})(1-x-2z-2y)]\cos\theta+[a_{0}\sqrt{1-q}(\frac{x}{2a_{1}}-a_{1})\\
&-q\sin2\zeta_{A}(\frac{1}{2}-\frac{1}{2}x-z-\frac{1}{2}a_{0}^{2}-y)]\sin\theta\cos\varphi-\frac{1}{2}x-\frac{1}{2}\},\\
F_{\max}^{11}=&\frac{2}{3}+\frac{1}{3}(-1+y+z),\\
F_{\max}^{12}=&\frac{2}{3}+\frac{1}{3}[-1+z+(2q\cos^{2}\zeta_{A}-1)y\cos\theta+qy\sin2\zeta_{A}\sin\theta\cos\varphi],\\
F_{\max}^{13}=&\frac{2}{3}+\frac{1}{3}\{[a_{0}^{2}(q\cos^{2}\zeta_{A}-1)+(\frac{1}{2}-q\cos^{2}\zeta_{A})(1+x-2z+2y)]\cos\theta+[-a_{0}\sqrt{1-q}(a_{1}+\frac{x}{2a_{1}})\\
&-q\sin2\zeta_{A}(\frac{1}{2}+\frac{1}{2}x-z-\frac{1}{2}a_{0}^{2}+y)]\sin\theta\cos\varphi+\frac{1}{2}x-\frac{1}{2}\},\\
F_{\max}^{14}=&\frac{2}{3}+\frac{1}{3}\{[a_{0}^{2}(q\cos^{2}\zeta_{A}-1)+(\frac{1}{2}-q\cos^{2}\zeta_{A})(1-x-2z)]\cos\theta+[a_{0}\sqrt{1-q}(\frac{x}{2a_{1}}-a_{1})\\
&-q\sin2\zeta_{A}(\frac{1}{2}-\frac{1}{2}x-z-\frac{1}{2}a_{0}^{2})]\sin\theta\cos\varphi-\frac{1}{2}x-y-\frac{1}{2}\},\\
F_{\max}^{15}=&\frac{2}{3}+\frac{1}{3}[-1+z+(1-2q\cos^{2}\zeta_{A})y\cos\theta-qy\sin2\zeta_{A}\sin\theta\cos\varphi],\\
F_{\max}^{16}=&\frac{2}{3}+\frac{1}{3}(-1-y+z).\\
\end{aligned}
\end{equation}

Eqs.(\ref{188}) and (\ref{189}) are an explicit expression of the maximal fidelity of the controlled teleportation. We can easily obtain the maximal fidelity when the measurement parameters $\theta, \varphi$,  resource state parameter $a_0$ and the noisy channels are determined.

In a given noisy channel,  if the controller Charlie is willing to help Alice and Bob to realize the teleportation, he can select the appropriate $\theta_{\rm m}$ and $\varphi_{\rm m}$, to maximize  ${F}_{\max}$. We use $\mathcal{F}_{\max}$ to stand for the maximum of the maximal average fidelity ${F}_{\max}$ and call it the  optimal average fidelity,  one can obtain  straightforwardly
\begin{equation}\label{888}
\mathcal{F}_{\max}={ \max}\{\frac{2}{3},\mathcal{F}_{\max}^{1}, \mathcal{F}_{\max}^{2},  \cdots, \mathcal{F}_{\max}^{16}\},
\end{equation}
where
\begin{equation}\label{288}\nonumber
\begin{aligned}
\mathcal{F}_{\max}^{1}=&\frac{2}{3}+\frac{1}{3}(x+y-z),~~\theta_{\rm m}~{\rm is}~{\rm arbitrary},~~\varphi_{\rm m}~{\rm is}~{\rm arbitrary};\\
\mathcal{F}_{\max}^{2}=&\mathcal{F}_{\max}^{5}=\frac{2-z}{3}+\frac{1}{3}\{[(2q\cos^{2}\zeta_{A}-1)(x+y)]^{2}+[\frac{a_{0}}{a_{1}}\sqrt{1-q}x+q(x+y)\sin2\zeta_{A}]^{2}\}^{\frac{1}{2}},\\
\theta_{\rm m}=&\frac{\pi}{2}-\arctan\frac{a_{1}(2q\cos^{2}\zeta_{A}-1)(x+y)}{a_{0}x\sqrt{1-q}+a_{1}q(x+y)\sin2\zeta_{A}},~~\varphi_{\rm m}=0~{\rm or}~2\pi;\\
\mathcal{F}_{\max}^{3}=&\mathcal{F}_{\max}^{9}=\frac{2}{3}+\frac{1}{3}\Big{\{}\{[a_{0}^{2}(1-q\cos^{2}\zeta_{A})+(q\cos^{2}\zeta_{A}-\frac{1}{2})(1+x-2z)]^{2}+[a_{0}\sqrt{1-q}(a_{1}+\frac{x}{2a_{1}})\\
&+q\sin2\zeta_{A}(\frac{1}{2}+\frac{1}{2}x-z-\frac{1}{2}a_{0}^{2})]^{2}\}^{\frac{1}{2}}+\frac{1}{2}x+y-\frac{1}{2}\Big{\}},\\
\theta_{\rm m}=&\frac{\pi}{2}-\arctan\frac{a_{0}^{2}(1-q\cos^{2}\zeta_{A})+(q\cos^{2}\zeta_{A}-\frac{1}{2})(1+x-2z)}{a_{0}\sqrt{1-q}(a_{1}+\frac{x}{2a_{1}})+q\sin2\zeta_{A}(\frac{1}{2}+\frac{1}{2}x-z-\frac{1}{2}a_{0}^{2})},~~\varphi_{\rm m}=0~{\rm or}~2\pi;\\
\end{aligned}
\end{equation}

\begin{equation}
\begin{aligned}
\mathcal{F}_{\max}^{4}=&\mathcal{F}_{\max}^{13}=\frac{2}{3}+\frac{1}{3}\Big{\{}\{[a_{0}^{2}(1-q\cos^{2}\zeta_{A})+(q\cos^{2}\zeta_{A}-\frac{1}{2})(1+x-2z+2y)]^{2}+[a_{0}\sqrt{1-q}(a_{1}+\frac{x}{2a_{1}})\\
                        &+q\sin2\zeta_{A}(\frac{1}{2}+\frac{1}{2}x-z-\frac{1}{2}a_{0}^{2}+y)]^{2}\}^{\frac{1}{2}}+\frac{1}{2}x-\frac{1}{2}\Big{\}},\\
\theta_{\rm m}=&\frac{\pi}{2}-\arctan\frac{a_{0}^{2}(1-q\cos^{2}\zeta_{A})+(q\cos^{2}\zeta_{A}-\frac{1}{2})(1+x-2z+2y)}{a_{0}\sqrt{1-q}(a_{1}+\frac{x}{2a_{1}})+q\sin2\zeta_{A}(\frac{1}{2}+\frac{1}{2}x-z-\frac{1}{2}a_{0}^{2}+y)},~~\varphi_{\rm m}=0~{\rm or}~2\pi;\\
\mathcal{F}_{\max}^{6}=&\frac{2}{3}-\frac{1}{3}(x+y+z),~~\theta_{\rm m}~{\rm is}~{\rm arbitrary},~~\varphi_{\rm m}~{\rm is}~{\rm arbitrary};\\
\mathcal{F}_{\max}^{7}=&\mathcal{F}_{\max}^{10}=\frac{2}{3}+\frac{1}{3}\Big{\{}\{[a_{0}^{2}(1-q\cos^{2}\zeta_{A})+(q\cos^{2}\zeta_{A}-\frac{1}{2})(1-x-2z-2y)]^{2}+[a_{0}\sqrt{1-q}(a_{1}-\frac{x}{2a_{1}})\\
&+q\sin2\zeta_{A}(\frac{1}{2}-\frac{1}{2}x-z-\frac{1}{2}a_{0}^{2}-y)]^{2}\}^{\frac{1}{2}}-\frac{1}{2}x-\frac{1}{2}\Big{\}},\\
\theta_{\rm m}=&\frac{\pi}{2}-\arctan\frac{a_{0}^{2}(1-q\cos^{2}\zeta_{A})+(q\cos^{2}\zeta_{A}-\frac{1}{2})(1-x-2z-2y)}{a_{0}\sqrt{1-q}(a_{1}-\frac{x}{2a_{1}})+q\sin2\zeta_{A}(\frac{1}{2}-\frac{1}{2}x-z-\frac{1}{2}a_{0}^{2}-y)},~~\varphi_{\rm m}=0~{\rm or}~2\pi;\\
\mathcal{F}_{\max}^{8}=&\mathcal{F}_{\max}^{14}=\frac{2}{3}+\frac{1}{3}\Big{\{}\{[a_{0}^{2}(1-q\cos^{2}\zeta_{A})+(q\cos^{2}\zeta_{A}-\frac{1}{2})(1-x-2z)]^{2}+[a_{0}\sqrt{1-q}(a_{1}-\frac{x}{2a_{1}})\\
&+q\sin2\zeta_{A}(\frac{1}{2}-\frac{1}{2}x-z-\frac{1}{2}a_{0}^{2})]^{2}\}^{\frac{1}{2}}-\frac{1}{2}x-y-\frac{1}{2}\Big{\}},\\
\theta_{\rm m}=&\frac{\pi}{2}-\arctan\frac{a_{0}^{2}(1-q\cos^{2}\zeta_{A})+(q\cos^{2}\zeta_{A}-\frac{1}{2})(1-x-2z)}{a_{0}\sqrt{1-q}(a_{1}-\frac{x}{2a_{1}})+q\sin2\zeta_{A}(\frac{1}{2}-\frac{1}{2}x-z-\frac{1}{2}a_{0}^{2})},~~\varphi_{\rm m}=0~{\rm or}~2\pi;\\
\mathcal{F}_{\max}^{11}=&\frac{2}{3}+\frac{1}{3}(-1+y+z),~~\theta_{\rm m}~{\rm is}~{\rm arbitrary},~~\varphi_{\rm m}~{\rm is}~{\rm arbitrary};\\
\mathcal{F}_{\max}^{12}=&\mathcal{F}_{\max}^{15}=\frac{2}{3}+\frac{1}{3}[\sqrt{[1-4q(1-q)\cos^{2}\zeta_{A}]y^{2}}-1+z],~~\theta_{\rm m}=\frac{\pi}{2}-\arctan\frac{-1+2q\cos^{2}\zeta_{A}}{q\sin2\zeta_{A}},~~\varphi_{\rm m}=0~{\rm or}~2\pi;\\
\mathcal{F}_{\max}^{16}=&\frac{2}{3}+\frac{1}{3}(-1-y+z),~~\theta_{\rm m}~{\rm is}~{\rm arbitrary},~~\varphi_{\rm m}~{\rm is}~{\rm arbitrary}.\\
\end{aligned}
\end{equation}

A detailed analysis, which is provided in Appendix \ref{A},  shows that
\begin{equation}\label{22}
\mathcal{F}_{\max}={\max}\{\frac {2}{3}, \mathcal{F}_{\max}^{1}, \mathcal{F}_{\max}^{2},   \mathcal{F}_{\max}^{3} , \mathcal{F}_{\max}^{7}\}
\end{equation}
for the case of $y\geq 0$; and
\begin{equation}\label{23}
\mathcal{F}_{\max}={\max}\{\frac {2}{3}, \mathcal{F}_{\max}^{1}, \mathcal{F}_{\max}^{2},   \mathcal{F}_{\max}^{4}, \mathcal{F}_{\max}^{6}, \mathcal{F}_{\max}^{8}\}
\end{equation}
for the case of $y\leq 0$.

Moreover,  we have  demonstrated that
\begin{equation}
\mathcal{F}_{\max}^{6}, ~~\mathcal{F}_{\max}^{7}, ~~ \mathcal{F}_{\max}^{8}\leq \frac{2}{3}.
\end{equation}
The detailed  proof can be referred to in  Appendix \ref{B}. As the classical threshold value of the maximal average fidelity of the teleportation without any entanglement is $\frac{2}{3}$, the optimal average fidelity must exceed the classical threshold.

Thus, the analytical expression for the  optimal average fidelity for $y>0$ was derived,
\begin{equation}
\begin{aligned}
\mathcal{F}_{\max}=\max\{\frac {2}{3}, \mathcal{F}_{\max}^{1},~\mathcal{F}_{\max}^{2},~\mathcal{F}_{\max}^{3}\}.
\end{aligned}
\end{equation}
Therefore, obtaining the optimal average fidelity in this case requires comparing the values and taking the maximum. Several values of the optimal average fidelity under different scenarios are listed in the table \ref{T1}.
\begin{table}[htbp]
\centering
\caption{The   optimal average fidelities  under different cases.}
\begin{tabular}{cccccccc}
\hline\hline
~~~~~~$a_{1}$~~~~~~ & ~~~~~~$q$~~~~~~ & ~~~~~~$k$~~~~~~ & ~~~~~~$r$~~~~~~ & ~~~~~~$\zeta_{A}$~~~~~~ & ~~~~~~$\zeta_{B}$~~~~~~ & ~~~~~~$\zeta_{C}$~~~~~~ & ~~~~~~$\mathcal{F}_{\max}$~~~~~~ \\
\hline
$0.5$ & $0.7$ & $0.8$ & $0.9$ & $0.78$ & $0.79$ & $0.8$ & $\mathcal{F}_{\max}^{1}=0.678821$ \\
$0.5$ & $0.3$ & $0.4$ & $0.5$ & $0.8$ & $0.9$ & $0.7$ & $\mathcal{F}_{\max}^{2}=0.790351$ \\
$0.5$ & $0.7$ & $0.8$ & $0.9$ & $0.5$ & $0.6$ & $0.7$ & $\mathcal{F}_{\max}^{3}=0.672405$ \\
$\frac{\sqrt 7}{4}$ & $0.7$ & $0.8$ & $0.9$ & $0.5$ & $0.6$ & $0.7$ & $\mathcal{F}_{\max}^{1}=0.682935$ \\
\hline\hline
\end{tabular}
\label{T1}
\end{table}

For the case of $y<0$, the  optimal average fidelity via the three noisy channels is
\begin{equation}
\begin{aligned}
\mathcal{F}_{\max}=\max\{\frac {2}{3}, \mathcal{F}_{\max}^{1},~\mathcal{F}_{\max}^{2},~\mathcal{F}_{\max}^{4}\}.
\end{aligned}
\end{equation}

It can be seen  that when the maximal average fidelity exceeds the classical threshold value $\frac {2}{3}$, its magnitude is related to the measurement angles chosen by Charlie. However, under generalized noisy channels, the loss of fidelity appears to depend on the channel parameters $\zeta_A, \zeta_B, \zeta_C$, the evolution parameters $q, k, r$, and the initial resource state parameter  $a_{1}$. Table \ref{T2} lists the  optimal average fidelities under different channel parameters and initial entangled state parameter $a_{1}$, which   indicates that for given the noisy channel parameters and initial resource state parameter $a_{1}$, the   optimal average fidelity can be calculated for all cases.

\begin{table}[htbp]
\caption{The optimal average fidelities under different noisy channel and resource state parameters.}
\centering
\begin{tabular}{cccccccc}
\hline
\hline
~~~~~~~~$a_{1}$~~~~~~~~ & ~~~~~~~~$q$~~~~~~~~ & ~~~~~~~~$k$~~~~~~~~ & ~~~~~~~~$r$~~~~~~~~ & ~~~~~~~~$\zeta_{A}$~~~~~~~~ & ~~~~~~~~$\zeta_{B}$~~~~~~~~ & ~~~~~~~~$\zeta_{C}$~~~~~~~~ & ~~~~~~~~$\mathcal{F}_{\max}$~~~~~~~~ \\
\hline
$0.5$ & $0.7$ & $0.8$ & $0.9$ & $0.78$ & $1$ & $4$ & $\mathcal{F}_{\max}^{1}=0.680900$ \\
$0.5$ & $0.2$ & $0.3$ & $0.4$ & $1.5$ & $1$ & $4$ & $\mathcal{F}_{\max}^{2}=0.843036$ \\
$0.5$ & $0.2$ & $0.8$ & $0.02$ & $\frac{\pi}{3}$ & $3.14$ & $1$ & $\mathcal{F}_{\max}^{4}=0.681082$ \\
$\frac{\sqrt{7}}{4}$ & $0.2$ & $0.8$ & $0.02$ & $\frac{\pi}{3}$ & $3.14$ & $1$ & $\mathcal{F}_{\max}^{4}=0.683770$ \\
\hline
\hline
\end{tabular}
\label{T2}
\end{table}

Although we have obtained a rigorous mathematical expression of the  optimal average fidelity, however since it contains seven parameters, one can not easily to see the visualized relationship between the  optimal average fidelity and these parameters.   In the following section, we exemplify in details the   optimal average fidelities in some special cases.

\section{The optimal average fidelity in some special cases}

 In this section we render some concrete scenarios  including  the absence of noisy channel,   the presence of one or two noisy channels,  as well as the case with three identical noisy channels to illustrate the properties of the optimal average fidelity.

\subsection{The optimal average fidelity in the standard controlled  teleportation}

In the standard controlled teleportation the noisy channel is  absence, that  corresponds to $\Lambda_{ABC}$ is an identity operator,  namely $q=k=r=0$. One can obtain the   optimal average fidelity in this case
\begin{equation}
\begin{aligned}
\mathcal{F}_{\max}=1,
\end{aligned}
\end{equation}
when Charlie chooses the measurement base decided   by   angles
\begin{equation}\label{29}
\begin{aligned}
\theta_{\rm m}=\frac{\pi}{2}+\arctan\frac{a_{1}}{a_{0}},~~\varphi_{\rm m}=0~{\rm or}~2\pi.
\end{aligned}
\end{equation}
 According to the Eq. (\ref{29}) it is clear that the selection of $\theta_{\rm m}$ is related to the ratio of $a_{1}$ to $a_{0}$. This case corresponds to the perfect controlled teleportation with unit fidelity.

\subsection{The optimal average fidelity through a noisy channel}

We now focus on the scenario where only qubit $A$ passes through the noisy quantum channel, which results in the emergence of the mixed resource state $\rho_{ABC}$ before the measurement phase of the protocol, i.e., the scenario under consideration corresponds to $|\Psi\rangle\langle\Psi|$ passing through a channel $\Lambda_{ABC}=\Lambda_{A}\otimes I\otimes I$, where $\Lambda_{A}$ encodes the interaction between qubit $A$ and its environment $E_{A}$. In this situation we have   $k=r=0$. One can easily to deduce that
 the   optimal average fidelity
\begin{equation}
\begin{aligned}
\mathcal{F}_{\max}=\max\{\mathcal{F}_{\max}^{1},~\mathcal{F}_{\max}^{2}\},
\end{aligned}
\end{equation}
where
\begin{equation}
\begin{aligned}
\mathcal{F}_{\max}^{1}&=\frac{2}{3}+\frac{\sqrt{2}a_{1}}{3},~~\theta_{\rm m}~{\rm is}~{\rm arbitrary},~~\varphi_{\rm m}~{\rm is}~{\rm arbitrary};\\
\mathcal{F}_{\max}^{2}&=\frac{2}{3}+\frac{\sqrt{2}a_{1}}{3}[(\frac{a_{0}}{a_{1}}\sqrt{1-q}+q\sin2\zeta_{A})^{2}+(1-2q\cos^{2}\zeta_{A})^{2}]^{\frac{1}{2}},\\
~\theta_{\rm m}&=\frac{\pi}{2}-\arctan\frac{2a_{1}q\cos^{2}\zeta_{A}-a_{1}}{a_{0}\sqrt{1-q}+a_{1}q\sin2\zeta_{A}},~\varphi_{\rm m}=0~{\rm or}~2\pi.\\
\end{aligned}
\end{equation}
\begin{figure}[htbp]
    \centering
    \begin{minipage}{0.33\textwidth}
       \centering
\includegraphics[width=\linewidth]{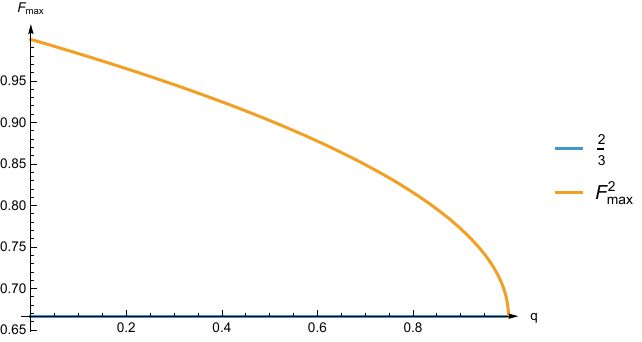}
        (a)
   \end{minipage}\hfill
   \begin{minipage}{0.33\textwidth}
       \centering
\includegraphics[width=\linewidth]{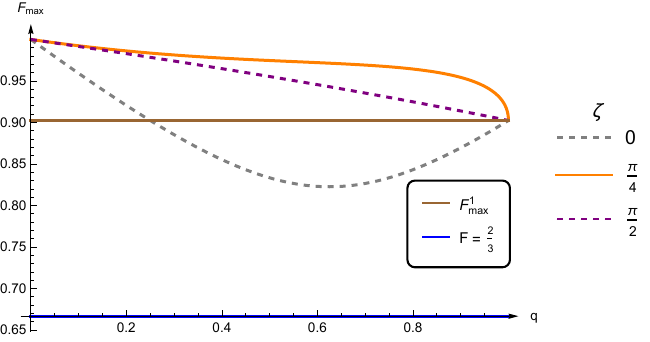}
       (b)
    \end{minipage}\hfill
    \begin{minipage}{0.33\textwidth}
        \centering
\includegraphics[width=\linewidth]{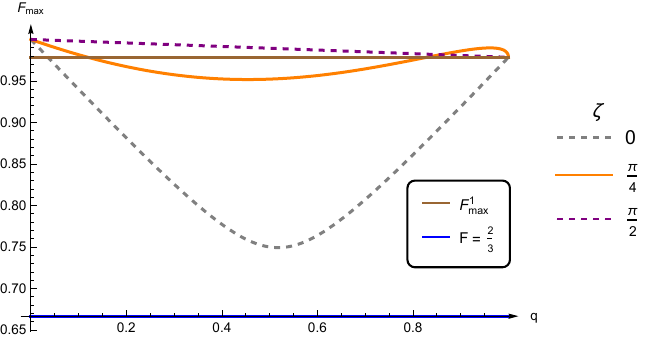}
       (c)
    \end{minipage}
  \caption{When $a_{1}=0$ (a), $a_{1}=\frac{1}{2}$ (b), $a_{1}=\frac{\sqrt{7}}{4}$ (c) the optimal average  fidelities of teleportation when qubit $A$ passes through a generalized noisy channel - exceeding the classical threshold value  $\frac{2}{3}$ - are  plotted as the function of the channel parameter $\zeta_{A}\in[0,2\pi]$ and $q$. By observing the influence of different $a_{1}$ on fidelity under various channels, we find that when $a_{1}<a_{0}$, the   optimal average fidelity decreases monotonically with increasing $q$. In contrast, when $a_{1}>a_{0}$, the    optimal average fidelity does not decrease monotonically with increasing $q$; instead, it first declines and then rises. This indicates that noisy channels do not always have a detrimental effect on fidelity.}
\label{F1}\end{figure}

Under a generalized noisy channel, the loss of fidelity depends on the channel parameter  $\zeta_{A}$, the evolution parameter $q$, and the initial resource state parameter $a_{1}$. Fixing the $\zeta_{A}$, when ${ a}_{1}>{ a}_{0}$, we find that the   optimal average fidelity first decreases and then increases as $q$ increases; when ${a}_{1}<{ a}_{0}$, the   optimal average fidelity decreases monotonically as $q$ increases.  The   optimal average fidelity can be seen from Fig.\ref{F1}, which plots the curves  $\mathcal{F}_{\max}^{1}$ and $\mathcal{F}_{\max}^{2}$. Clearly, in the left panel, the    optimal average fidelity is achieved by $\mathcal{F}_{\max}^{2}$. The middle and right panels also show the conditions that  when  $\mathcal{F}_{\max}^{1}$ is larger than $\mathcal{F}_{\max}^{2}$, or the contrary case. It can be easily observed that  in the real physical scenario of the controlled teleportation, the noisy channels do not always have a detrimental effect on fidelity.

Next, we analyse the scenario where only qubit $B$ passes through a noisy quantum channel. Then the  scenario considered corresponds to $|\Psi\rangle\langle\Psi|$ passing through a channel $\Lambda_{ABC}=I\otimes\Lambda_{B}\otimes I$, where $\Lambda_{B}$ encodes the interaction between qubit $B$ and his environment $E_{B}$. In this case,   $q=r=0$.
Without any difficult we derive that  the optimal average fidelity
\begin{equation}
\begin{aligned}
\mathcal{F}_{\max}=\max\{\frac{2}{3}, ~\mathcal{F}_{\max}^{2},~\mathcal{F}_{\max}^{3}\}, \\
\end{aligned}
\end{equation}
where
\begin{equation}
\begin{aligned}
\mathcal{F}_{\max}^{2}&=\frac{2}{3}+\frac{1}{3}(\sqrt{1-k}-\frac{1}{2}k\cos^{2}\zeta_{B}),\\
\mathcal{F}_{\max}^{3}&=\frac{2}{3}+\frac{1}{3}\{[(\frac{a_{0}}{\sqrt{2}}\sqrt{1-k}+a_{0}a_{1})^{2}+(\frac{1}{2}k\cos^{2}\zeta_{B}-\frac{1}{\sqrt{2}}a_{1}\sqrt{1-k}-a_{1}^{2})^{2}]^{\frac{1}{2}}+\frac{1}{\sqrt{2}}a_{1}\sqrt{1-k}-\frac{1}{2}\}.\\
\end{aligned}
\end{equation}
Here for the sake of simplicity, we omit the angles   $\theta_{\rm m}, \varphi_{\rm m}$.

\begin{figure}[h]
 \centering
\includegraphics[width=0.4\textwidth]{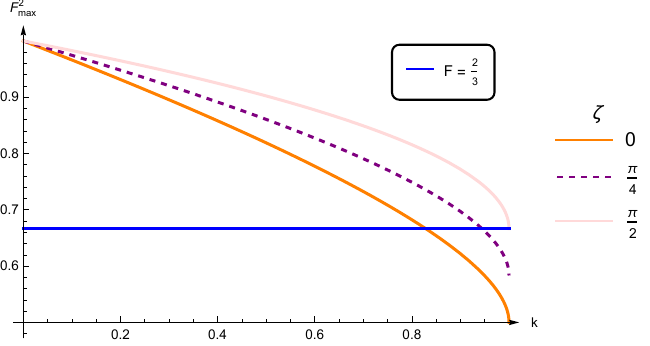}
  \caption{Evolutions of $\mathcal{F}_{\max}^{2}$ above the classical threshold value when the qubit $B$ is subject to the generalized noisy channels with different  channel parameters.}
\label{F2}
\end{figure}

We first analyze  $\mathcal{F}_{\max}^{2}$. A few typical cases are illustrated in  Fig.\ref{F2}. It is  found that $\mathcal{F}_{\max}^{2}$ decreases as the evolution parameter $k$ increases.

\begin{figure}[htbp]
    \centering
    \begin{minipage}{0.3\textwidth}
       \centering
\includegraphics[width=\linewidth]{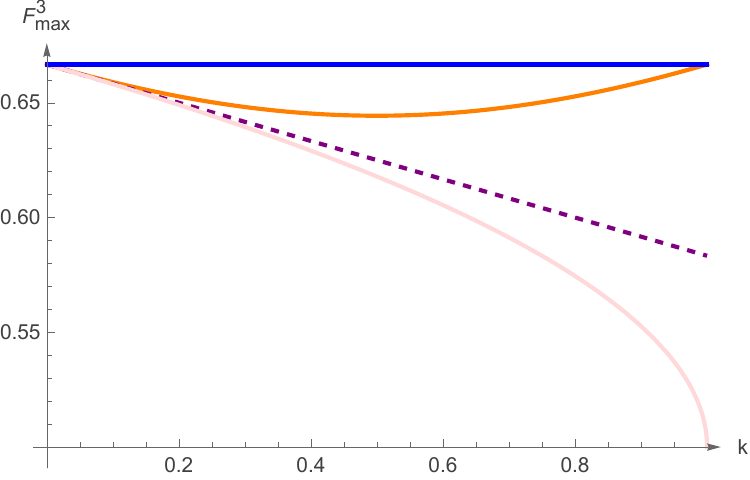}
        (a)
    \end{minipage}\hfill
    \begin{minipage}{0.3\textwidth}
       \centering
\includegraphics[width=\linewidth]{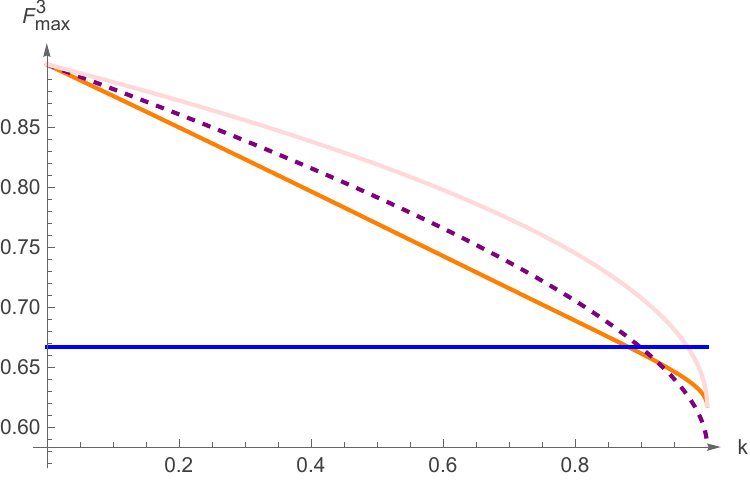}
       (b)
    \end{minipage}\hfill
    \begin{minipage}{0.35\textwidth}
        \centering
\includegraphics[width=\linewidth]{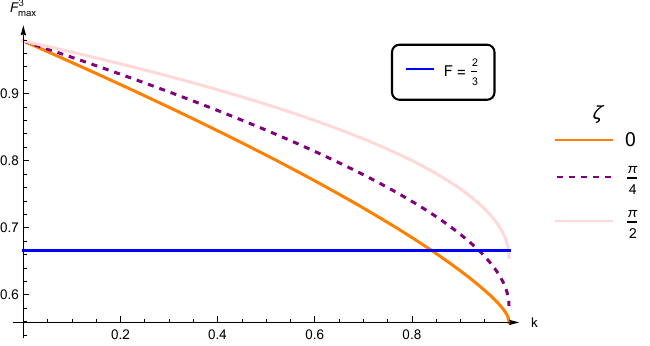}
       (c)
    \end{minipage}
  \caption{For $a_{1}=0$ (a), $a_{1}=\frac{1}{2}$ (b), $a_{1}=\frac{\sqrt{7}}{4}$ (c), $\mathcal{F}_{\max}^{3}$ are plotted against the channel parameter $\zeta_{B}$ and $k$, when qubit $B$ passes through a generalized noisy channel. Considering different channels, $\mathcal{F}_{\max}^{3}$ exhibits a decay as $\zeta_{B}$ increases from $0$ to $\frac{\pi}{2}$. Moreover, we find that increasing ${a}_{1}$ also helps to improve the fidelity.}
\label{F3}
\end{figure}

Next, we analyze  $\mathcal{F}_{\max}^{3}$, which is shown in Fig.\ref{F3}. Similarly, $\mathcal{F}_{\max}^{3}$ is found to decrease as the evolution parameter $k$ increases.

Based on the above fact, we find that increasing ${ a}_{1}$ also helps to improve the fidelity. When $a_{1}=0$, the maximum of $\mathcal{F}_{\max}^{3}$ can only reach the classical fidelity threshold value $\frac{2}{3}$ at most, and as $a_{1}$ gradually increases, $\mathcal{F}_{\max}^{3}$ is also   improved.

Owing to the symmetry between the second and third qubits, when qubit $C$ undergoing a noisy quantum channel yields a similar result.

\subsection{The optimal fidelity through two noisy channels}

We now consider the scenario where qubits $A$ and $B$ pass through the noisy channels. In this case, $|\Psi\rangle\langle\Psi|$ is subject to a  quantum channel $\Lambda_{ABC}=\Lambda_{A}\otimes \Lambda_{B}\otimes I$, while $\Lambda_{A}$ and $\Lambda_{B}$ represent the generalized noisy channels. For this special case, $r=0$, we have the optimal  average fidelity
\begin{equation}
\begin{aligned}
\mathcal{F}_{\max}=\max\{\frac {2}{3}, \mathcal{F}_{\max}^{1},~\mathcal{F}_{\max}^{2},~\mathcal{F}_{\max}^{3}\},
\end{aligned}
\end{equation}
 where
\begin{equation}
\begin{aligned}
\mathcal{F}_{\max}^{1}=&\frac{2}{3}+\frac{1}{3}(x-z),\\
\mathcal{F}_{\max}^{2}=&\frac{2}{3}+\frac{1}{3}\Big\{\{[(2q\cos^{2}\zeta_{A}-1)x]^{2}+[(\frac{a_{0}}{a_{1}}\sqrt{1-q}+q\sin2\zeta_{A})x]^{2}\}^{\frac{1}{2}}-z\Big\},\\
\mathcal{F}_{\max}^{3}=&\frac{2}{3}+\frac{1}{3}\Big\{\{[a_{0}^{2}(1-q\cos^{2}\zeta_{A})+(q\cos^{2}\zeta_{A}-\frac{1}{2})(1+x-2z)]^{2}+[a_{0}\sqrt{1-q}(a_{1}+\frac{x}{2a_{1}})\\
&+q\sin2\zeta_{A}(\frac{1}{2}+\frac{1}{2}x-z-\frac{1}{2}a_{0}^{2})]^{2}\}^{\frac{1}{2}}+\frac{1}{2}x-\frac{1}{2}\Big\}.\\
\end{aligned}
\end{equation}
Here
\begin{equation}
x=\sqrt{2}a_{1}\sqrt{1-k},      ~~~~ z=\frac{1}{2}k\cos^{2}\zeta_{B}.
\end{equation}

Next, we analyse the situation where qubits $B$ and $C$ pass through the noisy channels. In this case, the considered scenario corresponds to $|\Psi\rangle\langle\Psi|$ undergoing a channel $\Lambda_{ABC}=I\otimes\Lambda_{B}\otimes\Lambda_{C}$, where $\Lambda_{B}$ and $\Lambda_{C}$ are the generalized noisy channels. This case corresponds to $q=0$.

A direct calculation yields
\begin{equation}
\mathcal{F}_{\max}=\Big{\{}\begin{aligned}\max\{\frac{2}{3}, \mathcal{F}_{\max}^{1},~\mathcal{F}_{\max}^{2},~\mathcal{F}_{\max}^{3}\}, ~~~~{\rm if}  ~~~y\geq 0,\\
\max\{\frac{2}{3}, \mathcal{F}_{\max}^{1},~\mathcal{F}_{\max}^{2},~\mathcal{F}_{\max}^{4}\}, ~~~~{\rm if} ~~~~ y\leq 0,\\
\end{aligned}
\end{equation}
where
\begin{equation}
\begin{aligned}
\mathcal{F}_{\max}^{1}&=\frac{2}{3}+\frac{1}{3}(x+y-z),\\
\mathcal{F}_{\max}^{2}&=\frac{2-z}{3}+\frac{1}{3}\{[-(x+y)]^{2}+(\frac{a_{0}}{a_{1}}x)^{2}\}^{\frac{1}{2}},\\
\mathcal{F}_{\max}^{3}&=\frac{2}{3}+\frac{1}{3}\Big\{\{[a_{0}^{2}-\frac{1}{2}(1+x-2z)]^{2}+[a_{0}(a_{1}+\frac{x}{2a_{1}})]^{2}\}^{\frac{1}{2}}+\frac{1}{2}x+y-\frac{1}{2}\Big\},\\
\mathcal{F}_{\max}^{4}&=\frac{2}{3}+\frac{1}{3}\Big\{\{[a_{0}^{2}-\frac{1}{2}(1+x-2z+2y)]^{2}+[a_{0}(a_{1}+\frac{x}{2a_{1}})]^{2}\}^{\frac{1}{2}}+\frac{1}{2}x-\frac{1}{2}\Big\}.\\
\end{aligned}
\end{equation}
Here $x,y,z$ are defined by Eq. (\ref{xyz}).

\subsection{The optimal average fidelity through three identical channels}

Let us analyze a special case where qubits $A$, $B$ and $C$ all go through the same noisy channel, which is a usual case when three resource qubits in the same environment,  i.e., $q=k=r$ and $\zeta_{A}=\zeta_{B}=\zeta_{C}$. In this scenario, the optimal average fidelity reads
\begin{equation}
\begin{aligned}
\mathcal{F}_{\max}=\max\{\frac{2}{3},\mathcal{F}_{\max}^{1},~\mathcal{F}_{\max}^{2},~\mathcal{F}_{\max}^{3}\},
\end{aligned}
\end{equation}
where
\begin{equation}
\begin{aligned}
\mathcal{F}_{\max}^{1}&=\frac{2}{3}+\frac{1}{3}(x+y-z),\\
\mathcal{F}_{\max}^{2}&=\frac{2-z}{3}+\frac{1}{3}\{[(2q\cos^{2}\zeta_{A}-1)(x+y)]^{2}+[\frac{a_{0}}{a_{1}}\sqrt{1-q}x+q(x+y)\sin2\zeta_{A}]^{2}\}^{\frac{1}{2}},\\
\mathcal{F}_{\max}^{3}&=\frac{2}{3}+\frac{1}{3}\Big\{\{[a_{0}^{2}(1-q\cos^{2}\zeta_{A})+(q\cos^{2}\zeta_{A}-\frac{1}{2})(1+x-2z)]^{2}+[a_{0}\sqrt{1-q}(a_{1}+\frac{x}{2a_{1}})\\
&~~~~+q\sin2\zeta_{A}(\frac{1}{2}+\frac{1}{2}x-z-\frac{1}{2}a_{0}^{2})]^{2}\}^{\frac{1}{2}}+\frac{1}{2}x+y-\frac{1}{2}\Big\}\\
\end{aligned}
\end{equation}
with
\begin{equation}
\begin{aligned}
x=\sqrt{2}a_{1}(1-q), ~~~~y=\frac {1}{4}q^2\sin^{2}2\zeta_{A}, ~~~~z=q\cos^{2}\zeta_{A}(1-q\cos^{2}\zeta_{A}).
\end{aligned}
\end{equation}
In this scenario the optimal average fidelity is decided by three parameters $q$, $\zeta_{A}$ and $a_1$, its properties can therefore be readily analyzed.

\begin{figure}[htbp]
   \centering
   \begin{minipage}{0.3\textwidth}
       \centering
\includegraphics[width=\linewidth]{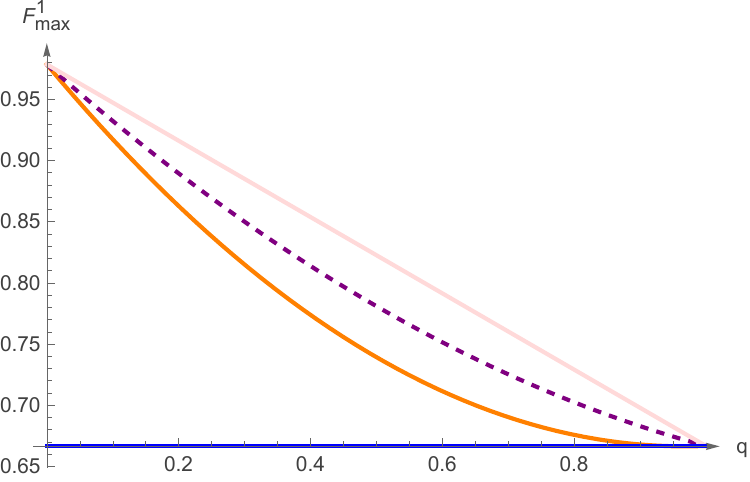}
        (a)
    \end{minipage}\hfill
    \begin{minipage}{0.3\textwidth}
        \centering
\includegraphics[width=\linewidth]{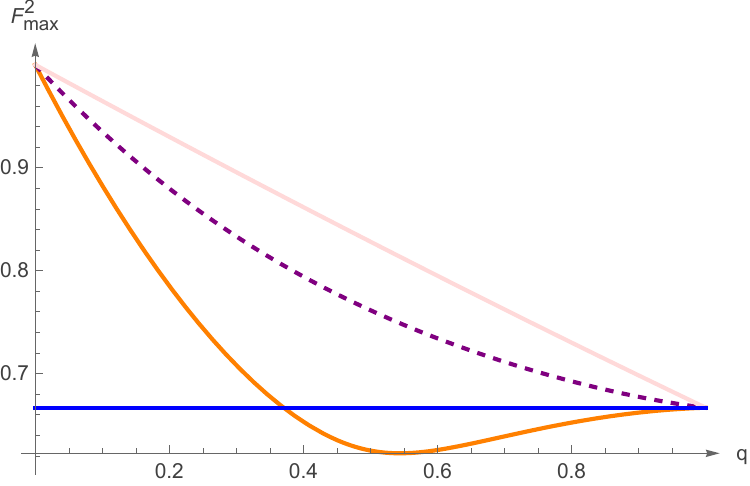}
       (b)
    \end{minipage}\hfill
   \begin{minipage}{0.35\textwidth}
       \centering
\includegraphics[width=\linewidth]{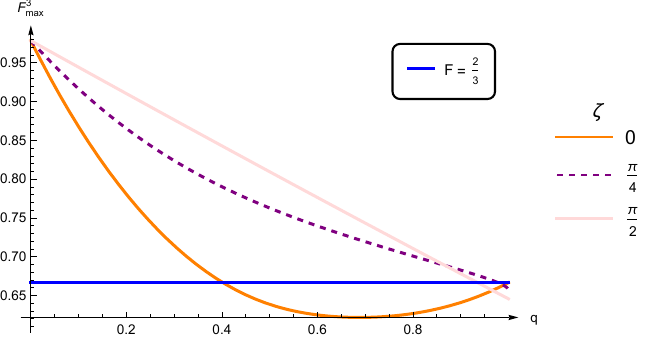}
        (c)
    \end{minipage}
  \caption{Evolution of (a) $\mathcal{F}_{\max}^{1}$, (b) $\mathcal{F}_{\max}^{2}$, and (c) $\mathcal{F}_{\max}^{3}$ for different values of the channel parameter $\zeta\in[0,2\pi]$, and $a_{1}=\frac{\sqrt{7}}{4}$. Obviously, the initial resource state is affected by the channels, resulting in a loss of the optimal average fidelity.}
\label{F4}
\end{figure}

Based on the above results, for this special case we illustrate the evolutions of $\mathcal{F}_{\max}^{1}$, $\mathcal{F}_{\max}^{2}$ and $\mathcal{F}_{\max}^{3}$ as a function of different channel parameters and $a_{1}=\frac{\sqrt{7}}{4}$ in Fig.\ref{F4}.

 The channels cause loss in the initial resource state, and the teleportation fidelity decreases accordingly. It can be clearly seen from Fig.\ref{F4} that the channel providing better fidelity is the dephasing channel, corresponding to $\zeta=\frac{\pi}{2}$. As shown in Figs.~\ref{F1}-\ref{F4}, the fidelity does not decay monotonically with the channel evolution parameter and this non-monotonic dependence arises from both the entanglement of the initial resource state and the channel parameters for the channel traversed by its first qubit.

\begin{figure}[h]
  \centering

\includegraphics[width=0.4\textwidth]{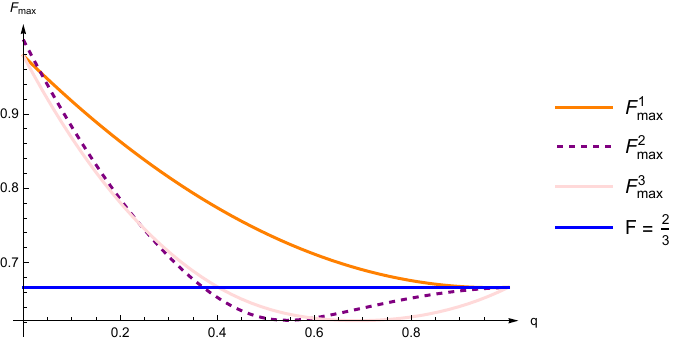}
  \caption{Evolution of the optimal average fidelity above the classical threshold value when the qubits $A$, $B$, $C$ are subject to the same generalized noisy channel, for  the channel parameter $\zeta_{A}=0$, resource state parameter $a_1=\frac{\sqrt 7}{4}$.}
\label{F5}
\end{figure}

 Therefore, the optimal  average fidelity can be obtained simply by extracting the fidelity curves for the same channel parameters from Fig.\ref{F4}. Interestingly, Fig.\ref{F5} presents such a figure, which is extracted from Fig.\ref{F4} for  $\zeta=0$. It can be clearly seen that when $q<0.045$, $\mathcal{F}_{\max}^{2}$ is the largest, and when $q>0.045$, $\mathcal{F}_{\max}^{1}$ is maximal. One can therefore directly read the optimal average fidelity for a given $q$ in this case from Fig.\ref{F5}.

\section{Conclusion}\label{Q8}

In summary, we study the controlled quantum teleportation in the presence of noisy channels. The impact of generalized noisy channels on the resource quantum state is systematically analysed. We define the maximal average fidelity and the optimal average fidelity, and derive their exact expressions for generalized noisy channels that continuously bridge the dephasing channels and amplitude damping channels, thereby encompassing extensive intermediate scenarios. To illustrate the influence of noisy channels on the optimal average fidelity of controlled quantum teleportation, we exemplify some special cases including the standard (noiseless) controlled quantum teleportation, a single qubit subjected to a noisy channel, two qubits each passing through noisy channels, and the scenario where all three qubits undergo the identical channel. It has been found that as the evolution parameters increase, the fidelity in some cases does not decay monotonically; instead, it first decreases and then increases. For this non-monotonic phenomenon, even when Charlie chooses a suitable measurement angle, the optimal fidelity still depends on seven parameters. We therefore restrict ourselves to special cases. Fortunately, by varying the degree of entanglement of the initial state, i.e., adjusting the ratio between $a_{0}$ and $a_{1}$, and by examining the scenarios in which different qubits of the initial resource state are subjected to a noisy channel, we obtained a set of plots. These reveal that the non-monotonic behavior becomes more pronounced as $a_{1}$ increases. Alternatively, with a fixed initial entanglement, the behavior differs depending on which qubit passes through the generalized noise channel. This property is most evident when the first qubit of the initial resource state traverses the generalized noise channel; when any of the other qubits traverses the channel, the fidelity exhibits only a monotonic decrease. That is in the real physical scenario of the controlled teleportation, the noisy channels do not always have a detrimental effect on fidelity.

As the fidelity  serves as a measure of teleportation success, and a reflection of  the real physical environmental noise, one can measure the fidelity to determine the noisy channel in the experiments.

\begin{acknowledgements}
This work was supported by
the National Natural Science Foundation of China under Grant No. 62271189.
\end{acknowledgements}

\appendix
\begin{widetext}

\section{An analysis of the optimal average  fidelity}\label{A}

By Eq. (\ref{288})  we obtain
\begin{equation}\label{A1}
\begin{aligned}
\mathcal{F}_{\max}^{1}-\mathcal{F}_{\max}^{6}&=\frac{2}{3}(x+y),\\
\mathcal{F}_{\max}^{1}-\mathcal{F}_{\max}^{11}&=\frac{1}{3}(1+x-2z),\\
\mathcal{F}_{\max}^{1}-\mathcal{F}_{\max}^{16}&=\frac{1}{3}(1+x-2z+2y),\\
\mathcal{F}_{\max}^{1}-\mathcal{F}_{\max}^{12}&=\frac{1}{3}(1+x-2z+y-\sqrt{[1-4q(1-q)\cos^2\zeta_A]y^{2}})\\
&\geq\frac{1}{3}(1+x-2z+y-\sqrt{y^{2}}).\\
\end{aligned}
\end{equation}
Let
$$\zeta=2\zeta_{A}, ~~~~\mu=2\zeta_{B}, ~~~~\nu=2\zeta_{C},$$
obviously,
\begin{equation}
\begin{aligned}
1-2z=\frac {1}{2}[(1-k)(1-r \cos\nu)+(1-r)(1-k\cos\mu)+kr(1+ \cos\mu\cos\nu)]\geq 0;\\
\end{aligned}
\end{equation}
\begin{equation}\label{A3}
\begin{aligned}
1-2z+2y&=\frac {1}{2}\{(1-k)(1-r \cos\nu)+(1-r)(1-k\cos\mu)+kr[1+ \cos(\mu-\nu)]\}\\
&\geq\frac {1}{2}\{(1-k)(1-r)+(1-r)(1-k)\}\\
&=(1-k)(1-r)\\
&\geq 0.\\
\end{aligned}
\end{equation}
So $\mathcal{F}_{\max}^{1}\geq \mathcal{F}_{\max}^{11}, \mathcal{F}_{\max}^{12}, \mathcal{F}_{\max}^{16}$. Thus  we can delete $\mathcal{F}_{\max}^{11}$, $\mathcal{F}_{\max}^{12}$, and $\mathcal{F}_{\max}^{16}$ in the Eq.(\ref{888}).

Next, let us compare $\mathcal{F}_{\max}^{3}$ and $\mathcal{F}_{\max}^{4}$. Let
$$s=1+x-2z, ~~~~t=a_{0}\sqrt{1-q}(2a_{1}+\frac{x}{a_{1}}).$$
 By Eq.(\ref{288}), we have
\begin{equation}\label{A4}
\begin{aligned}
6(\mathcal{F}_{\max}^{3}-\mathcal{F}_{\max}^{4})=&2y+\{[a_{0}^{2}(2-q-q\cos\zeta)+(q+q\cos\zeta-1)s]^{2}+[t+q\sin\zeta(s-a_{0}^{2})]^{2}\}^{\frac{1}{2}}\\
&-\{[a_{0}^{2}(2-q-q\cos\zeta)+(q+q\cos\zeta-1)(s+2y)]^{2}+[t+q\sin\zeta(s-a_{0}^{2}+2y)]^{2}\}^{\frac{1}{2}}.\\
\end{aligned}
\end{equation}
We will discuss $6(\mathcal{F}_{\max}^{3}-\mathcal{F}_{\max}^{4})$ in two cases: one is  $y\geq 0$, the other is $y\leq 0$.

1. When $y\geq 0$. In order to compare $\mathcal{F}_{\max}^{3}$ and $\mathcal{F}_{\max}^{4}$, according to Eq.(\ref{A4}) we should consider
\begin{equation}\label{A85}
\begin{aligned}
&~~~~{\big (}2y+\{[a_{0}^{2}(2-q-q\cos\zeta)+(q+q\cos\zeta-1)s]^{2}+[t+q\sin\zeta(s-a_{0}^{2})]^{2}\}^{\frac{1}{2}}{\big )}^{2}-{\big (}\{[a_{0}^{2}(2-q-q\cos\zeta)\\
&~~~~+(q+q\cos\zeta-1)(s+2y)]^{2}+[t+q\sin\zeta(s-a_{0}^{2}+2y)]^{2}\}^{\frac{1}{2}}{\big )}^{2}\\
&=4y\{4(1-q)qy\cos^{2}\frac{\zeta}{2}+W\},\\
\end{aligned}
\end{equation}
where
\begin{equation}
\begin{aligned}
W=&\{[a_{0}^{2}(2-q-q\cos\zeta)+(q+q\cos\zeta-1)s]^{2}+[t+q\sin\zeta(s-a_{0}^{2})]^{2}\}^{\frac{1}{2}}\\
&~~~~-[s(1+(-2q+2q^{2})(1+\cos\zeta))+qt\sin\zeta-(2-3q-3q\cos\zeta+2q^{2}+2q^{2}\cos\zeta)a_{0}^{2}].
\end{aligned}
\end{equation}
Let us  consider
\begin{equation}\label{quadric}
\begin{aligned}
&(\{[a_{0}^{2}(2-q-q\cos\zeta)+(q+q\cos\zeta-1)s]^{2}+[t+q\sin\zeta(s-a_{0}^{2})]^{2}\}^{\frac{1}{2}})^{2}-[s(1+(-2q+2q^{2})(1+\cos\zeta))\\
&+qt\sin\zeta-(2-3q-3q\cos\zeta+2q^{2}+2q^{2}\cos\zeta)a_{0}^{2}]^{2}\\
=&2q\cos^{2}\frac{\zeta}{2}(2-q-q\cos\zeta)x_{1}^{2}+(1-q^{2}\sin^{2}\zeta)y_{1}^{2}-2q(q+q\cos\zeta-1)\sin\zeta x_{1}y_{1},
\end{aligned}
\end{equation}
where
$$ x_{1}=a_{0}^{2}+(q+q\cos\zeta-1)(s-a_{0}^{2}),  ~~~~~y_{1}=t+q\sin\zeta(s-a_{0}^{2}).$$

Eq.(\ref{quadric}) is a quadric form about $x_1, y_1$. Because
\begin{equation}
\begin{aligned}
&2q\cos^{2}\frac{\zeta}{2}(2-q-q\cos\zeta)\geq 0,\\
&(1-q^{2}\sin^{2}\zeta)\geq 0,\\
&2q\cos^{2}\frac{\zeta}{2}(2-q-q\cos\zeta)(1-q^{2}\sin^{2}\zeta)-[q(q+q\cos\zeta-1)\sin\zeta]^{2}=4(1-q)q\cos^{2}\frac{\zeta}{2}\geq0,\\ \end{aligned}
\end{equation}
according to the criteria which determines that the quadric form is positive,
we have
\begin{equation}\label{A8}
\begin{aligned}
&(\{[a_{0}^{2}(2-q-q\cos\zeta)+(q+q\cos\zeta-1)s]^{2}+[t+q\sin\zeta(s-a_{0}^{2})]^{2}\}^{\frac{1}{2}})^{2}-[s(1+(-2q+2q^{2})(1+\cos\zeta))\\
&+qt\sin\zeta-(2-3q-3q\cos\zeta+2q^{2}+2q^{2}\cos\zeta)a_{0}^{2}]^{2}\geq0. \\
\end{aligned}
\end{equation}
Thus
\begin{equation}\label{A10}
W\geq 0.
\end{equation}
Combining Eqs.(\ref{A4}),  (\ref{A85}) and (\ref{A10}),  we arrive at the conclusion that when $y\geq 0$, then $\mathcal{F}_{\max}^{3}-\mathcal{F}_{\max}^{4}\geq0$.

2. When $y\leq 0$. By Eq.(\ref{A4}), if
\begin{equation}\label{A9}
2y+\{[a_{0}^{2}(2-q-q\cos\zeta)+(q+q\cos\zeta-1)s]^{2}+[t+q\sin\zeta(s-a_{0}^{2})]^{2}\}^{\frac{1}{2}}\leq 0,
\end{equation}
then
$\mathcal{F}_{\max}^{3}-\mathcal{F}_{\max}^{4}<0$.

Thus we only need to consider the case $0\geq y\geq -\frac {1}{2}\{[a_{0}^{2}(2-q-q\cos\zeta)+(q+q\cos\zeta-1)s]^{2}+[t+q\sin\zeta(s-a_{0}^{2})]^{2}\}^{\frac{1}{2}}.$ Obviously in this case
$2y+\{[a_{0}^{2}(2-q-q\cos\zeta)+(q+q\cos\zeta-1)s]^{2}+[t+q\sin\zeta(s-a_{0}^{2})]^{2}\}^{\frac{1}{2}}\geq 0$.

Now we consider
\begin{equation}\label{A5}
\begin{aligned}
&~~~~{\big (}2y+\{[a_{0}^{2}(2-q-q\cos\zeta)+(q+q\cos\zeta-1)s]^{2}+[t+q\sin\zeta(s-a_{0}^{2})]^{2}\}^{\frac{1}{2}}{\big )}^{2}-{\big (}\{[a_{0}^{2}(2-q-q\cos\zeta)\\
&~~~~+(q+q\cos\zeta-1)(s+2y)]^{2}+[t+q\sin\zeta(s-a_{0}^{2}+2y)]^{2}\}^{\frac{1}{2}}{\big )}^{2}\\
&=4y\{4(1-q)qy\cos^{2}\frac{\zeta}{2}+W\}.\\
\end{aligned}
\end{equation}
When $y=-\frac {1}{2}\{[a_{0}^{2}(2-q-q\cos\zeta)+(q+q\cos\zeta-1)s]^{2}+[t+q\sin\zeta(s-a_{0}^{2})]^{2}\}^{\frac{1}{2}}$  we have $$6(\mathcal{F}_{\max}^{3}-\mathcal{F}_{\max}^{4})=-\{[a_{0}^{2}(2-q-q\cos\zeta)+(q+q\cos\zeta-1)(s+2y)]^{2}+[t+q\sin\zeta(s-a_{0}^{2}+2y)]^{2}\}^{\frac{1}{2}}\leq 0.$$
That means
\begin{equation}
\begin{aligned}&4(1-q)qy\cos^{2}\frac{\zeta}{2}+W\geq 0
\end{aligned}\end{equation}
at $y=-\frac {1}{2}\{[a_{0}^{2}(2-q-q\cos\zeta)+(q+q\cos\zeta-1)s]^{2}+[t+q\sin\zeta(s-a_{0}^{2})]^{2}\}^{\frac{1}{2}}$.
By using Eq.(\ref{A10}), we obtain that  if $0\geq y\geq -\frac {1}{2}\{[a_{0}^{2}(2-q-q\cos\zeta)+(q+q\cos\zeta-1)s]^{2}+[t+q\sin\zeta(s-a_{0}^{2})]^{2}\}^{\frac{1}{2}}$, we must have
\begin{equation}\label{A12}
\begin{aligned}&4(1-q)qy\cos^{2}\frac{\zeta}{2}+W\geq 0.
\end{aligned}\end{equation}
That implies that when $0\geq y\geq -\frac {1}{2}\{[a_{0}^{2}(2-q-q\cos\zeta)+(q+q\cos\zeta-1)s]^{2}+[t+q\sin\zeta(s-a_{0}^{2})]^{2}\}^{\frac{1}{2}}$, we have $\mathcal{F}_{\max}^{3}-\mathcal{F}_{\max}^{4}\leq 0$.
By  Eqs. (\ref{A9}), (\ref{A5}) and  (\ref{A12}), one can easily deduce that when $y\leq 0$, then $\mathcal{F}_{\max}^{3}-\mathcal{F}_{\max}^{4}\leq 0$.

Similarly, one can obtain that if $y\geq 0$, then $\mathcal{F}_{\max}^{7}-\mathcal{F}_{\max}^{8}\geq 0$; when $y\leq 0$, then $\mathcal{F}_{\max}^{7}-\mathcal{F}_{\max}^{8}\leq 0$.

By using Eq.(\ref{A1}), one has if $y\geq 0$, then  $\mathcal{F}_{\max}^{1}\geq \mathcal{F}_{\max}^{6}.$ Therefore we arrive at the conclusion that Eqs. (\ref{22}) and (\ref{23})  hold.

\section{A proof of that $\mathcal{F}_{\max}^{6}$,  $\mathcal{F}_{\max}^{7}$,  and $\mathcal{F}_{\max}^{8}$ are less than the classical limit $\frac{2}{3}$}\label{B}

We will prove that $\mathcal{F}_{\max}^{6}$, $\mathcal{F}_{\max}^{7}$, and $\mathcal{F}_{\max}^{8}$ are less than the classical limit $\frac{2}{3}$ follows the order $\mathcal{F}_{\max}^{6}$, $\mathcal{F}_{\max}^{7}$, and $\mathcal{F}_{\max}^{8}$. By Eq.(\ref{288}), we obtain
\begin{equation}\label{B2}
\begin{aligned}
&~~~~~3(\frac{2}{3}-\mathcal{F}_{\max}^{6})\\
&=x+y+z\\
&=\sqrt{2}a_{1}\sqrt{(1-k)(1-r)}+\frac{1}{4}\{k[1+(1-r)\cos\mu]+r[1+(1-k)\cos\nu]-kr[1+\cos(\mu+\nu)]\}\\
&\geq\sqrt{2}a_{1}\sqrt{(1-k)(1-r)}+\frac{1}{4}\{k[1-(1-r)]+r[1-(1-k)]-2kr\}\\
&= \sqrt{2}a_{1}\sqrt{(1-k)(1-r)}\\
&\geq 0.\\
\end{aligned}
\end{equation}
It follows that $\mathcal{F}_{\max}^{6}\leq\frac{2}{3}$.

By Eq.(\ref{288}), we have
\begin{equation}
\begin{aligned}
6(\frac{2}{3}-\mathcal{F}_{\max}^{7})=&1+x-\{[a_{0}^{2}-(1-x-2y-2z-a_{0}^{2})(1-q-q\cos\zeta)]^{2}+[-q\sin\zeta(1-x-2y-2z-a_{0}^{2})\\
&+a_{0}\sqrt{1-q}(\sqrt{(2-2k)(1-r)}-2a_{1})]^{2}\}^{\frac{1}{2}}.\\
\end{aligned}
\end{equation}
Let
$$v=1-x-2y-2z-a_{0}^{2},$$
then we get
\begin{equation}
\begin{aligned}
6(\frac{2}{3}-\mathcal{F}_{\max}^{7})=&1+x-\{[a_{0}^{2}-v(1-q-q\cos\zeta)]^{2}+[-q v\sin\zeta+a_{0}\sqrt{1-q}(\sqrt{(2-2k)(1-r)}-2a_{1})]^{2}\}^{\frac{1}{2}}.\\
\end{aligned}
\end{equation}
It is easy to find the minimum value of $6(\frac{2}{3}-\mathcal{F}_{\max}^{7})$ respect to $\zeta$,
\begin{equation}
\begin{aligned}
(6(\frac{2}{3}-\mathcal{F}_{\max}^{7}))_{\rm mini}=1+x-q\sqrt{v^{2}}-\sqrt{[(q-1)v+a_{0}^{2}]^{2}-a_{0}^{2}(q-1)(\sqrt{(2-2k)(1-r)}-2a_{1})^{2}}.\\
\end{aligned}
\end{equation}

Now we discuss the bounds of $v$. One can deduce
\begin{equation}
\begin{aligned}
v=&1-x-2y-2z-a_{0}^{2}\\
=&\frac{1}{2}[2-k-r+kr+(k-1)r\cos\nu+(r-1)k\cos\mu+kr\cos(\mu+\nu)-2a_{0}^{2}-2a_{1}\sqrt{(2-2k)(1-r)}].\\
\end{aligned}
\end{equation}
Thus
\begin{equation}
\begin{aligned}
v\leq\frac{1}{2}[2-k-r+kr-(k-1)r-(r-1)k+kr-2a_{0}^{2}-2a_{1}\sqrt{(2-2k)(1-r)}]=1-a_{0}^{2}-a_{1}\sqrt{(2-2k)(1-r)}\leq1,\\
\end{aligned}
\end{equation}
\begin{equation}
\begin{aligned}
v&\geq\frac{1}{2}[2-k-r+kr+(k-1)r+(r-1)k-kr-2a_{0}^{2}-2a_{1}\sqrt{(2-2k)(1-r)}]\\
&=(k-1)(r-1)-a_{0}^{2}-a_{1}\sqrt{(2-2k)(1-r)}\\
&=(\sqrt{(1-k)(1-r)}-\frac{a_{1}}{\sqrt{2}})^{2}-a_{0}^{2}-\frac{a_{1}^{2}}{2}\\
&\geq-a_{0}^{2}-\frac{a_{1}^{2}}{2}\\
&=\frac{1}{4}(-1-2a_{0}^{2})\\
&\geq-\frac{1}{2}.\\
\end{aligned}
\end{equation}
Hence
  $$1+x-q\sqrt{v^{2}}\geq 0.$$

Next we  consider
\begin{equation}
\begin{aligned}
Y=&(1+x-q\sqrt{v^{2}})^{2}-(\sqrt{[(q-1)v+a_{0}^{2}]^{2}-a_{0}^{2}(q-1)(\sqrt{(2-2k)(1-r)}-2a_{1})^{2}})^{2}\\
=&1-v^{2}+2x+x^{2}+2v a_{0}^{2}-a_{0}^{4}+q(2v^{2}-2\sqrt{v^{2}}-2x\sqrt{v^{2}}-2va_{0}^{2}+a_{0}^{2}(\sqrt{(2-2k)(1-r)}-2a_{1})^{2})\\
&-a_{0}^{2}(\sqrt{(2-2k)(1-r)}-2a_{1})^{2}.\\
\end{aligned}
\end{equation}
Apparently, $Y$ is a linear function of $q$, the extrema of $Y$ respect to $q$ occur at the boundary points $q=0$ and $q=1$.

When $q=0$, we have
\begin{equation}
\begin{aligned}
Y&=1-v^{2}+2x+x^{2}+2va_{0}^{2}-a_{0}^{4}-a_{0}^{2}(\sqrt{(2-2k)(1-r)}-2a_{1})^{2}\\
&=(1+a_{1}\sqrt{(2-2k)(1-r)})^{2}-(v-a_{0}^{2})^{2}-a_{0}^{2}(\sqrt{(2-2k)(1-r)}-2a_{1})^{2}.\\
\end{aligned}
\end{equation}
It is observed that this is a quadratic function of a single variable $v$, with the maximal values of $Y$ occurring at the boundary of $v-a_{0}^{2}$.

When $v=1-a_{0}^{2}-a_{1}\sqrt{(2-2k)(1-r)}$, we have
\begin{equation}
\begin{aligned}
Y&=2(1-w^2)a_{0}^{2}+4w\sqrt{1-2a_{0}^{2}}\geq 0.\\
\end{aligned}
\end{equation}
Here $w=\sqrt{(1-k)(1-r)}$.

When $v=(k-1)(r-1)-a_{0}^{2}-a_{1}\sqrt{(2-2k)(1-r)}$, we get
\begin{equation}
Y=(1-w^2)(1+w^2-2a_{0}^{2})+2w(1+w^2)\sqrt{1-2a_{0}^{2}}\geq 0.
\end{equation}

Next, we analyze the situation when $q=1$. In this case we have
\begin{equation}
\begin{aligned}
Y=v^{2}-(2\sqrt{v^{2}}-1-x)(1+x)-a_{0}^{4}.\\
\end{aligned}
\end{equation}
If $v\geq0$, then
\begin{equation}
\begin{aligned}
Y=(1-v+x)^{2}-a_{0}^{4}.
\end{aligned}
\end{equation}
Thus the maximal values of $ Y$ occur at the boundary points $v=0$, $v=1-a_{0}^{2}-a_{1}\sqrt{(2-2k)(1-r)}$. When $v=0$, we have
\begin{equation}
\begin{aligned}
Y=(1+x)^{2}-a_{0}^{4}\geq 0.
\end{aligned}
\end{equation}
When $v=1-a_{0}^{2}-a_{1}\sqrt{(2-2k)(1-r)}$, we have
\begin{equation}
\begin{aligned}
Y&=(1-v+x)^{2}-a_{0}^{4}=(a_{0}^{2}+2w\sqrt{1-2a_{0}^{2}})^{2}-a_{0}^{4}\geq 0.\\
\end{aligned}
\end{equation}

If $v\leq0$, one  obtains
\begin{equation}
\begin{aligned}
Y=(1+v+x)^{2}-a_{0}^{4}.\\
\end{aligned}
\end{equation}

If $v=0$, one can deduce
\begin{equation}
\begin{aligned}
Y=(1+x)^{2}-a_{0}^{4}\geq0.
\end{aligned}
\end{equation}
When $v=(k-1)(r-1)-a_{0}^{2}-a_{1}\sqrt{(2-2k)(1-r)}$, we have
\begin{equation}
\begin{aligned}
Y&=(1+v+x)^{2}-a_{0}^{4}=(1+(k-1)(r-1)-a_{0}^{2})^{2}-a_{0}^{4}\geq (1-a_{0}^{2})^{2}-a_{0}^{4}=(1-2a_{0}^{2})\geq 0.\end{aligned}
\end{equation}
Thus $6(\frac{2}{3}-\mathcal{F}_{\max}^{7})\geq0$ always holds. Therefore we have demonstrated   $\mathcal{F}_{\max}^{7}\leq\frac{2}{3}$.

Next we use the same method to prove $\mathcal{F}_{\max}^{8}\leq\frac{2}{3}$. By  Eq.(\ref{288}), we get
\begin{equation}
\begin{aligned}
6(\frac{2}{3}-\mathcal{F}_{\max}^{8})=&1+x+2y-\{[a_{0}^{2}-(1-x-2z-a_{0}^{2})(1-q-q\cos\zeta)]^{2}+[-q\sin\zeta(1-x-2z-a_{0}^{2})\\
&+a_{0}\sqrt{1-q}(\sqrt{(2-2k)(1-r)}-2a_{1})]^{2}\}^{\frac{1}{2}}.\\
\end{aligned}
\end{equation}
Let
$$ u=1-x-2z-a_{0}^{2},$$
then one  derives
\begin{equation}
\begin{aligned}
6(\frac{2}{3}-\mathcal{F}_{\max}^{8})=&1+x+2y-\{[a_{0}^{2}-u(1-q-q\cos\zeta)]^{2}+[-qu\sin\zeta+a_{0}\sqrt{1-q}(\sqrt{(2-2k)(1-r)}-2a_{1})]^{2}\}^{\frac{1}{2}}.\\
\end{aligned}
\end{equation}
It is easy to find  the minimum value of $6(\frac{2}{3}-\mathcal{F}_{\max}^{8})$ respect to $\zeta$,
\begin{equation}\label{B21}
\begin{aligned}
(6(\frac{2}{3}-\mathcal{F}_{\max}^{8}))_{\rm mini}=1+x+2y-q\sqrt{u^{2}}-\sqrt{[(q-1)u+a_{0}^{2}]^{2}-a_{0}^{2}(q-1)(\sqrt{(2-2k)(1-r)}-2a_{1})^{2}}.\\
\end{aligned}
\end{equation}
Now we discuss the bounds of $u$. Apparently
\begin{equation}
\begin{aligned}
u=&1-x-2z-a_{0}^{2}\\
=&\frac{1}{4}[4-2k-2r+2kr+2k(r-1)\cos\mu+2r(k-1)\cos\nu+2kr\cos\mu\cos\nu]\\
&-a_{0}^{2}-\sqrt{2}a_{1}\sqrt{(1-k)(1-r)}.\\
\end{aligned}
\end{equation}
Thus
\begin{equation}
\begin{aligned}
u&\leq\frac{1}{4}[4-2k-2r+2kr-2(k-1)r-2(r-1)k+2kr-4a_{0}^{2}-4a_{1}\sqrt{(2-2k)(1-r)}]\\
&=1-a_{0}^{2}-a_{1}\sqrt{(2-2k)(1-r)}\\
&\leq1.\\
\end{aligned}
\end{equation}
\begin{equation}
\begin{aligned}
u&\geq\frac{1}{4}[4-2k-2r+2kr+2(k-1)r+2(r-1)k-2kr-4a_{0}^{2}-4a_{1}\sqrt{(2-2k)(1-r)}]\\
&=(k-1)(r-1)-a_{0}^{2}-a_{1}\sqrt{(2-2k)(1-r)}=(\sqrt{(1-k)(1-r)}-\frac{a_{1}}{\sqrt{2}})^{2}-a_{0}^{2}-\frac{a_{1}^{2}}{2}\\
&\geq-a_{0}^{2}-\frac{a_{1}^{2}}{2}\\
&=\frac{1}{4}(-1-2a_{0}^{2})\\
&\geq-\frac{1}{2}.\\
\end{aligned}
\end{equation}

Later on  we discuss  $1+x+2y-q\sqrt{u^{2}}$. Obviously,
\begin{equation}
1+x+2y-q\sqrt{u^{2}}\geq 1+x+2y-\sqrt{u^{2}}.
\end{equation}
If $u\geq 0$, we have
\begin{equation}
1+x+2y-\sqrt{u^{2}}= 1+x+2y-u=2(x+y+z)+a_{0}^{2}\geq 0,
\end{equation}
where we use $x+y+z\geq 0$, which has been demonstrated by Eq.(\ref{B2}).
If $u\leq 0$, one has
\begin{equation}
1+x+2y-\sqrt{u^{2}}= 1+x+2y+u=1-2z+2y+1-a_{0}^{2}\geq 0,
\end{equation}
where Eq. (\ref{A3})  indicated  $1-2z+2y\geq 0$ has been used.  Hence we have
$$1+x+2y-q\sqrt{u^{2}}\geq 0.$$

According to Eq.(\ref{B21}),  it is necessary to consider
\begin{equation}\label{B28}
\begin{aligned}
Z=&(1+x+2y-q\sqrt{u^{2}})^{2}-(\sqrt{[(q-1)u+a_{0}^{2}]^{2}-a_{0}^{2}(q-1)(\sqrt{(2-2k)(1-r)}-2a_{1})^{2}})^{2}\\
=&(1+x+2y)^{2}-u^{2}+2ua_{0}^{2}-a_{0}^{4}+q(2u^{2}-2\sqrt{u^{2}}-2x\sqrt{u^{2}}-2ua_{0}^{2}-4y\sqrt{u^{2}}+a_{0}^{2}(\sqrt{(2-2k)(1-r)}-2a_{1})^{2})\\
&-a_{0}^{2}(\sqrt{(2-2k)(1-r)}-2a_{1})^{2}.\\
\end{aligned}
\end{equation}
Evidenntly $Z$ is a linear function of $q$, the maximal values of $Z$ respect to $q$ occur at the boundary points $q=0$ and $q=1$. When $q=0$, we have
\begin{equation}
\begin{aligned}
Z=4(1+y-z-a_{0}^{2})(x+y+z+a_{0}^{2})-a_{0}^{2}(\sqrt{(2-2k)(1-r)}-2a_{1})^{2}.
\end{aligned}
\end{equation}

Let us analyze $1+y-z-a_{0}^{2}$. Apparently,
\begin{equation}
\begin{aligned}
1+y-z-a_{0}^{2}=\frac{1}{2}+y-z+\frac {1}{2}-a_{0}^{2}
\geq \frac{1}{2}+\frac{1}{2}(1-k)(1-r)-a_{0}^{2}
\geq 0.\\
\end{aligned}
\end{equation}
Here Eq.(\ref{A3}) has been used.
Then
\begin{equation}
\begin{aligned}
Z=&4(1+y-z-a_{0}^{2})(x+y+z+a_{0}^{2})-a_{0}^{2}(\sqrt{(2-2k)(1-r)}-2a_{1})^{2}\\
\geq&4(\frac{1}{2}+\frac{1}{2}(1-k)(1-r)-a_{0}^{2})a_{0}^{2}-a_{0}^{2}(\sqrt{(2-2k)(1-r)}-2a_{1})^{2}\\
=&4a_{0}^{2}a_{1}\sqrt{(2-2k)(1-r)}\\
\geq &0,\\
\end{aligned}
\end{equation}
where $x+y+z\geq0 $ mentioned  in Eq.(\ref{B2}) has been utilized. Therefore, in this situation, $Z\geq0$ holds.

Next, we analyze the case when $q=1$. In this situation Eq.(\ref{B28}) becomes
\begin{equation}
\begin{aligned}
Z=(1+x+2y-\sqrt{u^{2}})^{2}-a_{0}^{4}.\\
\end{aligned}
\end{equation}
If $u\geq0$, then
\begin{equation}
\begin{aligned}
Z=(1+x+2y-u)^{2}-a_{0}^{4}=[2(x+y+z)+a_{0}^{2}]^{2}-a_{0}^{4}\geq 0.
\end{aligned}
\end{equation}
If $u\leq 0$, we have
\begin{equation}
\begin{aligned}
Z=(1+x+2y+u)^{2}-a_{0}^{4}=(2+2y-2z-a_{0}^{2})^{2}-a_{0}^{4}=(1-2z+2y+\frac{1}{2}-a_{0}^{2}+\frac{1}{2})^{2}-a_{0}^{4}\geq \frac{1}{4}-a_{0}^{4}\geq  0.
\end{aligned}
\end{equation}
Here $1-2z+2y\geq 0$ demonstrated in Eq. (\ref{A3}), and  $a_{0}^{2}\leq\frac{1}{2}$, have been applied. Thus $6(\frac{2}{3}-\mathcal{F}_{\max}^{8})\geq0$ always holds, which induces $\mathcal{F}_{\max}^{8}\leq\frac{2}{3}$.

\end{widetext}

\end{document}